\documentclass[showpacs,showkeys,amssymb,onecolumn]{revtex4}
\usepackage{graphicx}
\usepackage{dcolumn}
\usepackage{multirow}
\usepackage{bm}
\usepackage{slashed}
\usepackage{enumerate}

\begin{document}

\title{$P$-wave charmed baryons from QCD sum rules}

\author{Hua-Xing Chen$^1$}
\email{hxchen@buaa.edu.cn}
\author{Wei Chen$^2$}
\email{wec053@mail.usask.ca}
\author{Qiang Mao$^{1,3}$}
\author{Atsushi Hosaka$^{4,5}$}
\email{hosaka@rcnp.osaka-u.ac.jp}
\author{Xiang Liu$^{6,7}$}
\email{xiangliu@lzu.edu.cn}
\author{Shi-Lin Zhu$^{8,9,10}$}
\email{zhusl@pku.edu.cn}
\affiliation{
$^1$School of Physics and Nuclear Energy Engineering and International Research Center for Nuclei and Particles in the Cosmos, Beihang University, Beijing 100191, China \\
$^2$Department of Physics and Engineering Physics, University of Saskatchewan, Saskatoon, Saskatchewan, S7N 5E2, Canada \\
$^3$Department of Electrical and Electronic Engineering, Suzhou University, Suzhou 234000, China\\
$^4$Research Center for Nuclear Physics, Osaka University, Ibaraki 567--0047, Japan \\
$^5$J-PARC Branch, KEK Theory Center, Institute of Particle and Nuclear Studies, KEK, Tokai, Ibaraki 319-1106, Japan \\
$^6$School of Physical Science and Technology, Lanzhou University, Lanzhou 730000, China\\
$^7$Research Center for Hadron and CSR Physics, Lanzhou University and Institute of Modern Physics of CAS, Lanzhou 730000, China\\
$^8$School of Physics and State Key Laboratory of Nuclear Physics and Technology, Peking University, Beijing 100871, China \\
$^9$Collaborative Innovation Center of Quantum Matter, Beijing 100871, China \\
$^{10}$Center of High Energy Physics, Peking University, Beijing 100871, China}

\begin{abstract}
We study the $P$-wave charmed baryons using the method of QCD sum rule in the framework of
heavy quark effective theory (HQET).
We consider systematically all possible baryon currents with a derivative for internal $\rho$- and $\lambda$-mode excitations. We have found a good working window for the currents corresponding to the $\rho$-mode excitations for $\Lambda_c(2595)$, $\Lambda_c(2625)$, $\Xi_c(2790)$, and $\Xi_c(2815)$ that complete two $SU(3)$ $\mathbf{\bar 3}_F$ multiplets of $J^P=1/2^-$ and $3/2^-$, while the currents corresponding to the $\lambda$-mode excitations seem also consistent with the data.
Our results also suggest that there are two $\Sigma_c(2800)$ states of $J^P = 1/2^-$ and $3/2^-$ whose mass splitting is $14 \pm 7$ MeV, and two $\Xi_c(2980)$ states whose mass splitting is $12 \pm 7$ MeV. They have two $\Omega_c$ partners of $J^P = 1/2^-$ and $3/2^-$, whose masses are around $3.25\pm0.20$ GeV with mass splitting $10 \pm 6$ MeV. All of them together complete two $SU(3)$ $\mathbf{6}_F$ multiplets of $J^P=1/2^-$ and $3/2^-$. They may also have $J^P=5/2^-$ partners. $\Xi_c(3080)$ may be one of them, and the other two are $\Sigma_c(5/2^-)$ and $\Omega_c(5/2^-)$, whose masses are $85 \pm 23$ MeV and $50 \pm 27$ MeV larger.
\end{abstract}

\pacs{14.20.Lq, 12.38.Lg, 12.39.Hg}
\keywords{excite heavy baryons, QCD sum rule, heavy quark effective theory}
\maketitle

\section{Introduction}

In the past years important progress has been made in the field of heavy baryons.
All the ground state charmed baryons containing a single charm quark have been well established
both experimentally and theoretically~\cite{pdg}. The lowest-lying orbitally excited charmed states
$\Lambda_c(2595)$ ($J^P=1/2^-$), $\Lambda_c(2625)$ ($J^P=3/2^-$), $\Xi_c(2790)$ ($J^P=1/2^-$) and $\Xi_c(2815)$ ($J^P=3/2^-$)
have been well observed by several collaborations and complete two $SU(4)$ $\mathbf{\bar4}$
multiplets~\cite{pdg,Albrecht:1993pt,Frabetti:1993hg,Edwards:1994ar,Alexander:1999ud}.
Besides them, several $P$-wave charm baryon candidates $\Sigma_c(2800)$ ($J^P=?^?$), $\Xi_c(2980)$ ($J^P=?^?$) and $\Xi_c(3080)$ ($J^P=?^?$) are also well observed by the
Belle and BaBar collaborations~\cite{Mizuk:2004yu,Aubert:2008ax,Chistov:2006zj,Aubert:2007dt},
and more data are expected in the near future.

These heavy baryons are also interesting in a theoretical point of view~\cite{Korner:1994nh,Manohar:2000dt,Klempt:2009pi}.
The light degrees of freedom (quarks and gluons) circle around the nearly static heavy
quark, and the whole system behaves as the QCD analogue of
the familiar hydrogen bounded by electromagnetic interaction.
In the past two decades, various phenomenological
models have been used to study heavy baryons, including the relativized potential quark model~\cite{Capstick:1986bm}, the
Feynman-Hellmann theorem~\cite{Roncaglia:1995az}, the combined expansion in $1/m_Q$ and $1/N_c$~\cite{Jenkins:1996de},
the relativistic quark model~\cite{Ebert:2007nw}, the chiral quark model~\cite{Zhong:2007gp}, the hyperfine interaction~\cite{Copley:1979wj,Karliner:2008sv}, the pion induced reactions~\cite{Kim:2014qha},
the variational approach~\cite{Roberts:2007ni}, the Faddeev approach~\cite{Garcilazo:2007eh}, the constituent quark model~\cite{Ortega:2012cx},
the unitarized dynamical model~\cite{GarciaRecio:2012db}, the extended local hidden gauge approach~\cite{Liang:2014eba}, the unitarized chiral perturbation theory~\cite{Lu:2014ina}, etc.
There are also many Lattice QCD studies~\cite{Bowler:1996ws,Burch:2008qx}; see a recent reference for more details~\cite{Brown:2014ena}.

In this paper we shall study the
$P$-wave charmed baryons using the method of QCD sum rule~\cite{Shifman:1978bx,Reinders:1984sr} in the
framework of the heavy quark effective theory
(HQET)~\cite{Grinstein:1990mj,Eichten:1989zv,Falk:1990yz}. This method has
been successful for studying heavy mesons containing a single heavy
quark, as done in Refs.~\cite{Bagan:1991sg,Neubert:1991sp,Neubert:1993mb,Broadhurst:1991fc,Ball:1993xv,Huang:1994zj,Dai:1996yw,Dai:1993kt,Dai:1996qx,Colangelo:1998ga,Dai:2003yg}.
Recently, we applied it to study $D$-wave $\bar c s$ heavy mesons in Ref.~\cite{Zhou:2014ytp}.
This method has also been successful for studying heavy baryons containing a single heavy
quark, as done in Refs.~\cite{Shuryak:1981fza,Grozin:1992td,Bagan:1993ii,Dai:1995bc,Dai:1996xv,Groote:1996em,Zhu:2000py,Lee:2000tb,Huang:2000tn,Wang:2003zp}.
Particularly, we have applied this method to systematically study the ground state bottom baryons
in Ref.~\cite{Liu:2007fg}, and we found that the extracted chromomagnetic
splitting between the bottom baryon heavy doublet agrees well with the experimental data.
We note that some studies using the method of QCD sum rules not in HQET but
in full QCD can be found in Refs.~\cite{Bagan:1992tp,Bagan:1991sc,Duraes:2007te,Wang:2007sqa}.
In this paper we shall follow the procedures used in
these references, and systematically study the $P$-wave charmed baryons. We shall also
consider the ${\mathcal O}(1/m_Q)$ corrections and extract the chromomagnetic splitting,
with $m_Q$ being the heavy quark mass.

This paper is organized as follows. In Sec.~\ref{sec:current}, we
systematically construct the interpolating currents for the $P$-wave charmed baryons. Then
in Sec.~\ref{sec:leading} we use one of them as an example to
perform the QCD sum rule analysis at the leading order, and in
Sec.~\ref{sec:nexttoleading} we still use it as an example to perform the QCD sum rule analysis by taking into account the ${\mathcal O}(1/m_Q)$
corrections. In
Sec.~\ref{sec:summary} we summarize and discuss our results.

\section{interpolating fields for the $P$-wave charmed baryon}
\label{sec:current}

The $P$-wave charmed baryons have been systemically classified in Ref.~\cite{Chen:2007xf}, where the strong decays of heavy baryons were investigated systematically using the $^3P_0$ model. In this paper we classify the $P$-wave charmed baryon interpolating fields using the same notations, i.e., $l_\rho$ denotes the orbital angular momentum between the two light quarks and $l_\lambda$ denotes the orbital angular momentum between the charm quark and the two-light-quark system. To describe these structures using interpolating fields, we can use either local fields or those containing derivatives~\cite{Zhu:2000py,Lee:2000tb,Huang:2000tn,Wang:2003zp}. In this paper we use the latter ones, because they can describe the inner structures of charmed baryons in a more clear way.

Generally, the interpolating field for charmed baryons can be written as a combination of a diquark field and a heavy quark field£º
\begin{eqnarray}
J(x) \sim \epsilon_{abc} \left( q^{aT}(x) \mathbb{C} \Gamma_1 q^b (x) \right) \Gamma_2 h_v^c (x) \, .
\label{eq:baryonfield}
\end{eqnarray}
We note that the derivatives are not explicitly shown in this equation. $a$, $b$ and $c$ are color indices, and $\epsilon_{abc}$ is the totally antisymmetric tensor; the superscript $T$ represents the transpose of the Dirac indices only; $\mathbb{C}$ is the charge-conjugation operator. $q(x)$ is the light quark field at location $x$, and it can be either $u(x)$ or $d(x)$ or $s(x)$. $h_v(x)$ is the heavy quark field, and we have used the Fierz transformation to move it to the rightmost place.

\begin{figure}[htb]
\begin{center}
\scalebox{0.6}{\includegraphics{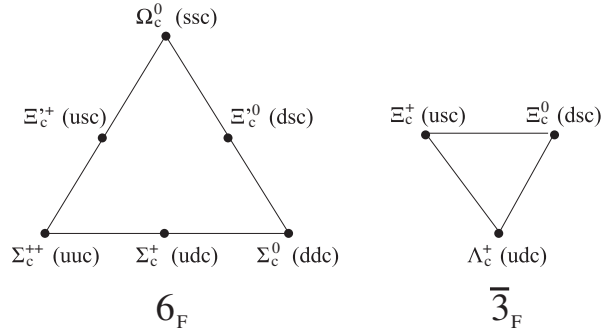}}
\end{center}
\caption{The $SU(3)$ flavor multiplets of charmed baryons.
\label{fig:baryon}}
\end{figure}

There are two ``good'' $S$-wave diquark fields. One is
\begin{eqnarray}
\epsilon_{abc} q^{aT}(x) \mathbb{C} \gamma_5 q^b(x) \, , \, \,  [^1S_0] \, ,
\end{eqnarray}
which has quantum numbers $j_l^{P_l} = 0^+$. It has orbital angular momentum $l_\rho=0$, so its orbital degree of freedom is symmetric ($\mathbf{S}$); it has spin angular momentum $s_l=0$, so its spin degree of freedom is antisymmetric ($\mathbf{A}$); it has the antisymmetric color structure $\mathbf{\bar 3}_C$ ($\mathbf{A}$). Therefore, it should have the antisymmetric flavor structure $\mathbf{\bar 3}_F$  ($\mathbf{A}$) due to the Pauli principle, although this is not shown explicitly (see the right panel of Fig.~\ref{fig:baryon}). The other $S$-wave diquark field is
\begin{eqnarray}
\epsilon_{abc} q^{aT}(x) \mathbb{C} \gamma_\mu q^b(x) \, , \, \, [^3S_1] \, ,
\end{eqnarray}
which has quantum numbers $j_l^{P_l} = 1^+$, $l_\rho=0$ ($\mathbf{S}$), $s_l=1$ ($\mathbf{S}$), color $\mathbf{\bar 3}_C$ ($\mathbf{A}$) and flavor $\mathbf{6}_F$ ($\mathbf{S}$) (see the left panel of Fig.~\ref{fig:baryon}).

\begin{figure}[htb]
\begin{center}
\scalebox{0.6}{\includegraphics{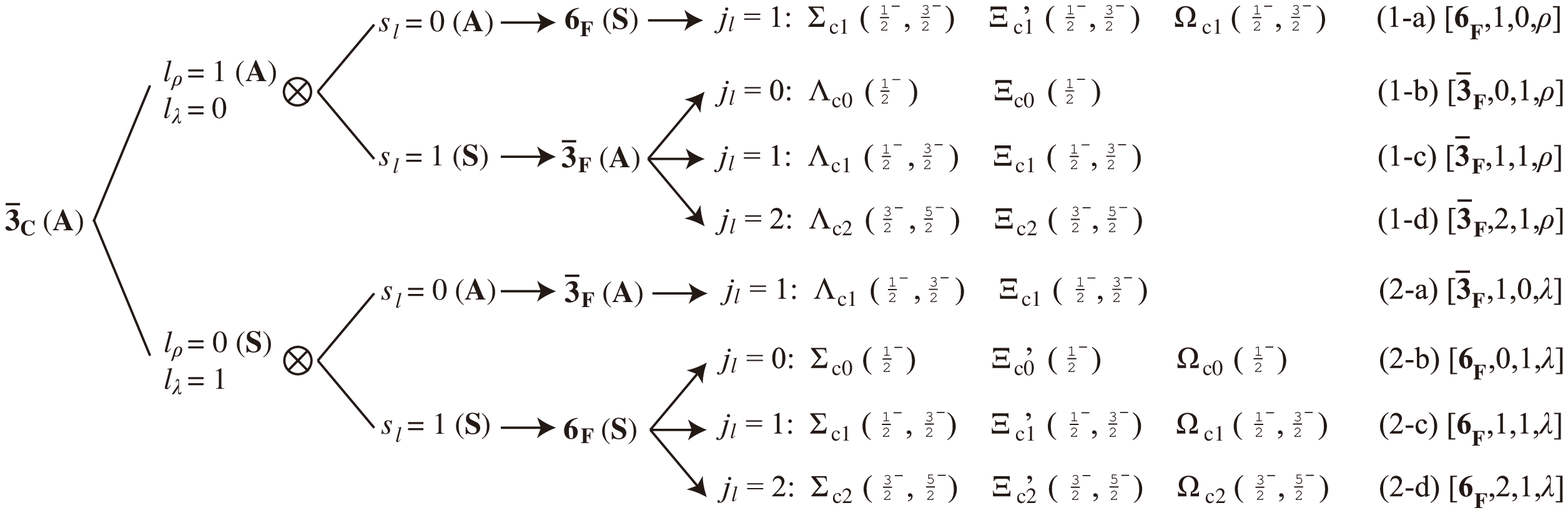}}
\end{center}
\caption{The notations for $P$-wave charmed baryons. $\mathbf{6}_F$ ($\mathbf{S}$) and $\mathbf{\bar 3}_F$ ($\mathbf{A}$) denote the $SU(3)$ flavor representations. $\mathbf{\bar 3}_C$ ($\mathbf{A}$) denotes the $SU(3)$ color representation. $s_l$ is the spin angular momentum of the two light quarks, and $j_l = l_\lambda \otimes l_\rho \otimes s_l$ is the total angular momentum of the two light quarks.
\label{fig:pwave}}
\end{figure}

The $P$-wave diquark fields can be obtained by adding a derivative to these $S$-wave diquark fields, either between the two light quarks [$l_\rho=1$ ($\mathbf{A}$) and $l_\lambda=0$], or between the charm quark and the two-light-quark system [$l_\rho=0$ ($\mathbf{S}$) and $l_\lambda=1$]:
\begin{eqnarray}
&& \epsilon_{abc} \Big ( [\mathcal{D}^{\nu} q^{aT}(x)] \mathbb{C} \gamma_5 q^b(x) - q^{aT}(x) \mathbb{C} \gamma_5 [\mathcal{D}^{\nu} q^b(x)] \Big ) \, , \, \,  [^3P_0] \, , \, l_\rho=1~(\mathbf{A}) \, , \, l_\lambda=0 \, ,
\\ && \epsilon_{abc} \Big ( [\mathcal{D}^{\nu} q^{aT}(x)] \mathbb{C} \gamma_\mu q^b(x) - q^{aT}(x) \mathbb{C} \gamma_\mu [\mathcal{D}^{\nu} q^b(x)] \Big ) \, , \, \, [^1P_1]/[^3P_0]/[^5P_1] \, , \, l_\rho=1~(\mathbf{A}) \, , \, l_\lambda=0 \, ,
\\ && \epsilon_{abc} \Big ( [\mathcal{D}^{\nu} q^{aT}(x)] \mathbb{C} \gamma_5 q^b(x) + q^{aT}(x) \mathbb{C} \gamma_5 [\mathcal{D}^{\nu} q^b(x)] \Big ) \, , \, \,  [^1S_0] \, , \, l_\rho=0~(\mathbf{S}) \, , \, l_\lambda=1 \, ,
\\ && \epsilon_{abc} \Big ( [\mathcal{D}^{\nu} q^{aT}(x)] \mathbb{C} \gamma_\mu q^b(x) + q^{aT}(x) \mathbb{C} \gamma_\mu [\mathcal{D}^{\nu} q^b(x)] \Big ) \, , \, \, [^3S_1] \, , \, l_\rho=0~(\mathbf{S}) \, , \, l_\lambda=1 \, ,
\end{eqnarray}
where $D^\mu = \partial^\mu - i g A^\mu$ is the gauge-covariant derivative. They can be further used to construct the $P$-wave charmed baryon fields $J^{\alpha_1\cdots\alpha_{j-1/2}}_{j,P,F,j_l,s_l,\rho/\lambda}$, where $j$, $P$, and $F$ denote the total angular momentum, parity and $SU(3)$ flavor representation ($\mathbf{\bar 3}_F$ or $\mathbf{6}_F$) of the charmed baryons; $j_l$ and $s_l$ denote the total angular momentum and spin angular momentum of the light components; $\rho$ denotes $l_\rho = 1$ and $l_\lambda = 0$, while $\lambda$ denotes $l_\rho = 0$ and $l_\lambda = 1$. We have the relations $j_l = l_\lambda \otimes l_\rho \otimes s_l$ and $j = j_l \otimes s_Q$, where $s_Q = 1/2$ is the spin of the heavy quark. The results are summarized in Fig.~\ref{fig:pwave}, while their explicit forms are the following:
\begin{enumerate}[(1)]

\item $l_\rho = 1$ ($\mathbf{A}$) and $l_\lambda = 0$:

\begin{enumerate}[({1}-a)]

\item $s_l=0$ ($\mathbf{A}$). We denote this case as $[\mathbf{6}_F, 1, 0, \rho]$. Now the diquark has quantum numbers $j_l = l_\lambda \otimes l_\rho \otimes s_l = 1$, color representation $\mathbf{\bar 3}_C$ ($\mathbf{A}$), and flavor representation $\mathbf{6}_F$ ($\mathbf{S}$). The total angular momentum of the charm baryon is $j = j_l \otimes s_Q = 1/2 \oplus 3/2$, so we obtain a doublet $(j^P = 1/2^- , 3/2^-)$:
\begin{eqnarray}
J_{1/2,-,\mathbf{6}_F,1,0,\rho} &=& i \epsilon_{abc} \Big ( [\mathcal{D}_t^{\mu} q^{aT}] C \gamma_5 q^b -  q^{aT} C \gamma_5 [\mathcal{D}_t^{\mu} q^b] \Big ) \gamma_t^{\mu} \gamma_5 h_v^c \, ,
\label{eq:current1}
\\ J^{\alpha}_{3/2,-,\mathbf{6}_F,1,0,\rho} &=& i \epsilon_{abc} \Big ( [\mathcal{D}_t^{\mu} q^{aT}] C \gamma_5 q^b -  q^{aT} C \gamma_5 [\mathcal{D}_t^{\mu} q^b] \Big ) \Big ( g_t^{\alpha\mu} - {1\over3} \gamma_t^{\alpha} \gamma_t^{\mu} \Big ) h_v^c \, ,
\label{eq:current2}
\end{eqnarray}
where $\gamma_t^\mu = \gamma^\mu - v\!\!\!\slash v^\mu$, $D^\mu_t = D^\mu - (D \cdot v) v^\mu$,  $h_v$ is the heavy quark field in HQET, $v$ is the velocity of the heavy quark, and $g_t^{\alpha_1\alpha_2}=g^{\alpha_1\alpha_2} - v^{\alpha_1} v^{\alpha_2}$ is the transverse metric tensor.

\item $s_l = 1$ ($\mathbf{S}$) and $j_l = 0$. We denote this case as $[\mathbf{\bar 3}_F, 0, 1, \rho]$. Now the diquark has color $\mathbf{\bar 3}_C$ ($\mathbf{A}$) and flavor $\mathbf{\bar 3}_F$ ($\mathbf{A}$), and we obtain a baryon singlet $(1/2^-)$:
\begin{eqnarray}
J_{1/2,-,\mathbf{\bar3}_F,0,1,\rho} &=& i \epsilon_{abc} \Big ( [\mathcal{D}_t^{\mu} q^{aT}] C \gamma_t^\mu q^b -  q^{aT} C \gamma_t^\mu [\mathcal{D}_t^{\mu} q^b] \Big ) h_v^c \, .
\label{eq:current3}
\end{eqnarray}

\item $s_l = 1$ ($\mathbf{S}$) and $j_l = 1$. We denote this case as $[\mathbf{\bar 3}_F, 1, 1, \rho]$. Now the diquark has color $\mathbf{\bar 3}_C$ ($\mathbf{A}$), and flavor $\mathbf{\bar 3}_F$ ($\mathbf{A}$), and we obtain a baryon doublet $(1/2^- , 3/2^-)$:
\begin{eqnarray}
J_{1/2,-,\mathbf{\bar3}_F,1,1,\rho} &=& i \epsilon_{abc} \Big ( [\mathcal{D}_t^{\mu} q^{aT}] C \gamma_t^\nu q^b -  q^{aT} C \gamma_t^\nu [\mathcal{D}_t^{\mu} q^b] \Big ) \sigma_t^{\mu\nu} h_v^c \, ,
\label{eq:current4}
\\ J^{\alpha}_{3/2,-,\mathbf{\bar3}_F,1,1,\rho} &=&
\label{eq:current5}
i \epsilon_{abc} \Big ( [\mathcal{D}_t^{\mu} q^{aT}] C \gamma_t^\nu q^b - q^{aT} C \gamma_t^\nu [\mathcal{D}_t^{\mu} q^b] \Big )
\\ \nonumber && \times \Big ( g_t^{\alpha\mu} \gamma_t^{\nu} \gamma_5 - g_t^{\alpha\nu} \gamma_t^{\mu} \gamma_5 - {1 \over 3} \gamma_t^{\alpha} \gamma_t^{\mu} \gamma_t^{\nu} \gamma_5 + {1 \over 3} \gamma_t^{\alpha} \gamma_t^{\nu} \gamma_t^{\mu} \gamma_5 \Big ) h_v^c \, .
\end{eqnarray}

\item $s_l = 1$ ($\mathbf{S}$) and $j_l = 2$. We denote this case as $[\mathbf{\bar 3}_F, 2, 1, \rho]$. Now the diquark has color $\mathbf{\bar 3}_C$ ($\mathbf{A}$) and flavor $\mathbf{\bar 3}_F$ ($\mathbf{A}$), and we obtain a baryon doublet $(3/2^- , 5/2^-)$:
\begin{eqnarray}
J^{\alpha}_{3/2,-,\mathbf{\bar3}_F,2,1,\rho} &=&
\label{eq:current6}
i \epsilon_{abc} \Big ( [\mathcal{D}_t^{\mu} q^{aT}] C \gamma_t^\nu q^b - q^{aT} C \gamma_t^\nu [\mathcal{D}_t^{\mu} q^b] \Big )
\times \Big ( g_t^{\alpha\mu} \gamma_t^{\nu} \gamma_5 + g_t^{\alpha\nu} \gamma_t^{\mu} \gamma_5 - {2 \over 3} g_t^{\mu\nu} \gamma_t^{\alpha} \gamma_5 \Big ) h_v^c \, ,
\\ J^{\alpha_1\alpha_2}_{5/2,-,\mathbf{\bar3}_F,2,1,\rho} &=& i \epsilon_{abc} \Big ( [\mathcal{D}_t^{\alpha_1} q^{aT}] C \gamma_t^{\alpha_2} q^b - q^{aT} C \gamma_t^{\alpha_2} [\mathcal{D}_t^{\alpha_1} q^b] + [\mathcal{D}_t^{\alpha_2} q^{aT}] C \gamma_t^{\alpha_1} q^b - q^{aT} C \gamma_t^{\alpha_1} [\mathcal{D}_t^{\alpha_2} q^b]
\label{eq:current7}
\\ \nonumber && - {2\over3} g_t^{\alpha_1\alpha_2} g_t^{\mu \nu} \times \big( [\mathcal{D}_t^{\mu} q^{aT}] C \gamma_t^{\nu} q^b - q^{aT} C \gamma_t^{\nu} [\mathcal{D}_t^{\mu} q^b] \big ) \Big ) h_v^c \, .
\end{eqnarray}

\end{enumerate}

\item $l_\rho = 0$ ($\mathbf{S}$) and $l_\lambda = 1$:

\begin{enumerate}[({2}-a)]

\item $s_l=0$ ($\mathbf{A}$). We denote this case as $[\mathbf{\bar 3}_F, 1, 0, \lambda]$. Now the diquark has quantum numbers $j_l=1$, color $\mathbf{\bar 3}_C$ ($\mathbf{A}$) and flavor $\mathbf{\bar 3}_F$ ($\mathbf{A}$), and we obtain a baryon doublet $(1/2^- , 3/2^-)$:
\begin{eqnarray}
J_{1/2,-,\mathbf{\bar3}_F,1,0,\lambda} &=& i \epsilon_{abc} \Big ( [\mathcal{D}_t^{\mu} q^{aT}] C \gamma_5 q^b +  q^{aT} C \gamma_5 [\mathcal{D}_t^{\mu} q^b] \Big ) \gamma_t^{\mu} \gamma_5 h_v^c \, ,
\label{eq:current8}
\\ J^{\alpha}_{3/2,-,\mathbf{\bar3}_F,1,0,\lambda} &=& i \epsilon_{abc} \Big ( [\mathcal{D}_t^{\mu} q^{aT}] C \gamma_5 q^b +  q^{aT} C \gamma_5 [\mathcal{D}_t^{\mu} q^b] \Big ) \Big ( g_t^{\alpha\mu} - {1\over3} \gamma_t^{\alpha} \gamma_t^{\mu} \Big ) h_v^c \, .
\label{eq:current9}
\end{eqnarray}

\item $s_l = 1$ ($\mathbf{S}$) and $j_l = 0$. We denote this case as $[\mathbf{6}_F, 0, 1, \lambda]$. Now the diquark has color $\mathbf{\bar 3}_C$ ($\mathbf{A}$) and flavor $\mathbf{6}_F$ ($\mathbf{S}$), and we obtain a baryon singlet $(1/2^-)$:
\begin{eqnarray}
J_{1/2,-,\mathbf{6}_F,0,1,\lambda} &=& i \epsilon_{abc} \Big ( [\mathcal{D}_t^{\mu} q^{aT}] C \gamma_t^\mu q^b +  q^{aT} C \gamma_t^\mu [\mathcal{D}_t^{\mu} q^b] \Big ) h_v^c \, .
\label{eq:current10}
\end{eqnarray}

\item $s_l = 1$ ($\mathbf{S}$) and $j_l = 1$. We denote this case as $[\mathbf{6}_F, 1, 1, \lambda]$. Now the diquark has color $\mathbf{\bar 3}_C$ ($\mathbf{A}$) and flavor $\mathbf{6}_F$ ($\mathbf{S}$), and we obtain a baryon doublet $(1/2^- , 3/2^-)$:
\begin{eqnarray}
J_{1/2,-,\mathbf{6}_F,1,1,\lambda} &=& i \epsilon_{abc} \Big ( [\mathcal{D}_t^{\mu} q^{aT}] C \gamma_t^\nu q^b +  q^{aT} C \gamma_t^\nu [\mathcal{D}_t^{\mu} q^b] \Big ) \sigma_t^{\mu\nu} h_v^c \, ,
\label{eq:current11}
\\ J^{\alpha}_{3/2,-,\mathbf{6}_F,1,1,\lambda} &=&
\label{eq:current12}
i \epsilon_{abc} \Big ( [\mathcal{D}_t^{\mu} q^{aT}] C \gamma_t^\nu q^b + q^{aT} C \gamma_t^\nu [\mathcal{D}_t^{\mu} q^b] \Big )
\\ \nonumber && \times \Big ( g_t^{\alpha\mu} \gamma_t^{\nu} \gamma_5 - g_t^{\alpha\nu} \gamma_t^{\mu} \gamma_5 - {1 \over 3} \gamma_t^{\alpha} \gamma_t^{\mu} \gamma_t^{\nu} \gamma_5 + {1 \over 3} \gamma_t^{\alpha} \gamma_t^{\nu} \gamma_t^{\mu} \gamma_5 \Big ) h_v^c \, .
\end{eqnarray}

\item $s_l = 1$ ($\mathbf{S}$) and $j_l = 2$. We denote this case as $[\mathbf{6}_F, 2, 1, \lambda]$. Now the diquark has color $\mathbf{\bar 3}_C$ ($\mathbf{A}$) and flavor $\mathbf{6}_F$ ($\mathbf{S}$), and we obtain a baryon doublet $(3/2^- , 5/2^-)$:
\begin{eqnarray}
J^{\alpha}_{3/2,-,\mathbf{6}_F,2,1,\lambda} &=&
\label{eq:current13}
i \epsilon_{abc} \Big ( [\mathcal{D}_t^{\mu} q^{aT}] C \gamma_t^\nu q^b + q^{aT} C \gamma_t^\nu [\mathcal{D}_t^{\mu} q^b] \Big )
\times \Big ( g_t^{\alpha\mu} \gamma_t^{\nu} \gamma_5 + g_t^{\alpha\nu} \gamma_t^{\mu} \gamma_5 - {2 \over 3} g_t^{\mu\nu} \gamma_t^{\alpha} \gamma_5 \Big ) h_v^c \, ,
\\ J^{\alpha_1\alpha_2}_{5/2,-,\mathbf{6}_F,2,1,\lambda} &=& i \epsilon_{abc} \Big ( [\mathcal{D}_t^{\alpha_1} q^{aT}] C \gamma_t^{\alpha_2} q^b + q^{aT} C \gamma_t^{\alpha_2} [\mathcal{D}_t^{\alpha_1} q^b] + [\mathcal{D}_t^{\alpha_2} q^{aT}] C \gamma_t^{\alpha_1} q^b + q^{aT} C \gamma_t^{\alpha_1} [\mathcal{D}_t^{\alpha_2} q^b]
\label{eq:current14}
\\ \nonumber && - {2\over3} g_t^{\alpha_1\alpha_2} g_t^{\mu \nu} \times \big( [\mathcal{D}_t^{\mu} q^{aT}] C \gamma_t^{\nu} q^b + q^{aT} C \gamma_t^{\nu} [\mathcal{D}_t^{\mu} q^b] \big ) \Big ) h_v^c \, .
\end{eqnarray}

\end{enumerate}

\end{enumerate}
We note that all these interpolating fields have been projected to have either $j=1/2$ or $j=3/2$, except $J^{\alpha_1\alpha_2}_{5/2,-,\mathbf{\bar3}_F,2,1,\rho}$ and $J^{\alpha_1\alpha_2}_{5/2,-,\mathbf{6}_F,2,1,\lambda}$, which contain both $j=3/2$ and $j=5/2$ components.

Identical sum rules can be obtained using either $J_{1/2,-,\mathbf{6}_F,1,0,\rho}$ or $J^{\alpha}_{3/2,-,\mathbf{6}_F,1,0,\rho}$ in the same doublet,
both at the leading order and at the $O(1/m_Q)$ order~\cite{Dai:1993kt,Dai:1996yw,Dai:1996qx,Dai:2003yg}.
We note that actually there are small and negligible differences. So do other currents in the same doublet.
Accordingly, we do not need to use all of them to perform QCD sum rule
analyses. In this paper we use $J_{1/2,-,\mathbf{6}_F,1,0,\rho}$, $J_{1/2,-,\mathbf{\bar3}_F,0,1,\rho}$,
$J_{1/2,-,\mathbf{\bar3}_F,1,1,\rho}$, $J^{\alpha}_{3/2,-,\mathbf{\bar3}_F,2,1,\rho}$, $J_{1/2,-,\mathbf{\bar3}_F,1,0,\lambda}$,
$J_{1/2,-,\mathbf{6}_F,0,1,\lambda}$, $J_{1/2,-,\mathbf{6}_F,1,1,\lambda}$ and $J^{\alpha}_{3/2,-,\mathbf{6}_F,2,1,\lambda}$ to
perform QCD sum rule analyses and
study the baryon multiplets $[\mathbf{6}_F,1,0,\rho]$, $[\mathbf{\bar3}_F,0,1,\rho]$, $[\mathbf{\bar3}_F,1,1,\rho]$, $[\mathbf{\bar3}_F,2,1,\rho]$, $[\mathbf{\bar3}_F,1,0,\lambda]$, $[\mathbf{6}_F,0,1,\lambda]$, $[\mathbf{6}_F,1,1,\lambda]$ and $[\mathbf{6}_F,2,1,\lambda]$, respectively.

Before performing sum rule analyses, we need to explicitly write out the quark contents contained in these currents. This can be easily done
according to Fig.~\ref{fig:baryon}. We use similar symbols to denote them based on previous symbols $J^{\alpha_1\cdots\alpha_{j-1/2}}_{j,P,F,j_l,s_l,\rho/\lambda}(x)$ and $[F,j_l,s_l,\rho/\lambda]$, just with $\mathbf{6}_F$ replaced by $\Sigma_c$, $\Xi_c^\prime$, and $\Omega_c$, and $\mathbf{\bar3}_F$ replaced by $\Lambda_c$ and $\Xi_c$. For example, $J_{1/2,-,\Xi_c,1,1,\rho}$ is used to denote
$J_{1/2,-,\mathbf{\bar3}_F,1,1,\rho}$ with quark contents $usc$ (or $dsc$):
\begin{eqnarray}
J_{1/2,-,\Xi_c,1,1,\rho} &=& i \epsilon_{abc} \Big ( [\mathcal{D}_t^{\mu} u^{aT}] C \gamma_t^\nu s^b -  u^{aT} C \gamma_t^\nu [\mathcal{D}_t^{\mu} s^b] \Big ) \sigma_t^{\mu\nu} h_v^c \, .
\label{eq:current15}
\end{eqnarray}
In the following sections we use this current as an example to perform the QCD sum rule analysis and study the baryon doublet $[\Xi_c,1,1,\rho]$ containing $\Xi_c({1/2}^-)$ and $\Xi_c({3/2}^-)$.

\section{The Sum Rules at the Leading Order (in the $m_Q \rightarrow \infty$ limit)}
\label{sec:leading}

In the previous section we have classified the $P$-wave charmed baryon interpolating fields, and in this and the next sections we use them to perform QCD sum rule analyses. When classifying these fields, we have taken into account their inner structures by fixing their inner quantum numbers $j_l$, $s_l$, $l_\rho$, and $l_\lambda$. However, the physical state may be a mixed state containing components of different inner quantum numbers. If this is the case, different currents can well couple to the same physical states. For example, the observed $\Sigma_c(2800)$ state ($J^P = 1/2^-$) may contain both a $\rho$ component ($l_\rho = 1$ and $l_\lambda = 0$) and $\lambda$ component ($l_\rho = 0$ and $l_\lambda = 1$), and then the two currents $J_{1/2,-,\Sigma_c,1,0,\rho}$ and $J_{1/2,-,\Sigma_c,1,0,\lambda}$ may both well couple to it.

We keep this in mind, but at the beginning we can always assume different currents couple to different states. We use
$|j,P,F,j_l,s_l,\rho/\lambda\rangle$ to denote the heavy baryon state with the quantum
numbers $j$, $P$, $F$ and the inner quantum numbers $j_l$, $s_l$, and $\rho/\lambda$
in the $m_Q \rightarrow \infty$ limit, and assume that the relation between this state and
the relevant interpolating field is
\begin{eqnarray}
\langle 0| J_{1/2,P,F,j_l,s_l,\rho/\lambda}(x) |1/2,P,F,j_l,s_l,\rho/\lambda \rangle
= f_{1/2,P,F,j_l,s_l,\rho/\lambda} u(x) \, ,
\\
\langle 0| J^\alpha_{3/2,P,F,j_l,s_l,\rho/\lambda}(x) |3/2,P,F,j_l,s_l,\rho/\lambda \rangle
= f_{3/2,P,F,j_l,s_l,\rho/\lambda} u^\alpha(x) \, .
\end{eqnarray}
In these equations $f_{j,P,F,j_l,s_l,\rho/\lambda}$ is used to denote the decay constant, which has the same value for
the two states in the same doublet in the $m_Q \rightarrow \infty$ limit. $u(x)$ and $u^\alpha(x)$ are the Dirac and Rarita-Schwinger
spinors, respectively. These currents can be used to construct the two-point correlation function
\begin{eqnarray}
\Pi^{\alpha_1\cdots\alpha_{j-1/2},\beta_1\cdots\beta_{j-1/2}}_{j,P,F,j_l,s_l,\rho/\lambda} (\omega)
&=& i \int d^4 x e^{i k x} \langle 0 |
T[J^{\alpha_1\cdots\alpha_{j-1/2}}_{j,P,F,j_l,s_l,\rho/\lambda}(x)
\bar J^{\beta_1\cdots\beta_{j-1/2}}_{j,P,F,j_l,s_l,\rho/\lambda}(0)] | 0 \rangle\nonumber
\\  &=& \mathbb{S} [ g_t^{\alpha_1 \beta_1} \cdots g_t^{\alpha_{j-1/2} \beta_{j-1/2}} ] \Pi_{j,P,F,j_l,s_l,\rho/\lambda} (\omega) \, .\label{eq:pi}
\end{eqnarray}
In this equation $\omega$ is used to denote twice the external off-shell energy, $\omega = 2 v \cdot k$,
and $\mathbb{S}[\cdots]$ is used to denote symmetrization and subtracting the trace
terms in the sets $(\alpha_1 \cdots \alpha_{j-1/2})$ and $(\beta_1 \cdots
\beta_{j-1/2})$. We can write Eq.~(\ref{eq:pi}) at the hadron level as
\begin{eqnarray}
\Pi_{j,P,F,j_l,s_l,\rho/\lambda}(\omega) = {f_{j,P,F,j_l,s_l,\rho/\lambda}^{2} \over 2 \overline{\Lambda}_{j,P,F,j_l,s_l,\rho/\lambda}
- \omega} + \mbox{higher states} \, . \label{eq:pole}
\end{eqnarray}
In this equation $\overline{\Lambda}_{j,P,F,j_l,s_l,\rho/\lambda}$ is defined to be
\begin{eqnarray}
\overline{\Lambda}_{j,P,F,j_l,s_l,\rho/\lambda} \equiv \lim_{m_Q \rightarrow \infty} (m_{j,P,F,j_l,s_l,\rho/\lambda} - m_Q) \, ,
\end{eqnarray}
where $m_{j,P,F,j_l,s_l,\rho/\lambda}$ is used to denote the mass of the lowest-lying heavy baryon state to which
$J^{\alpha_1\cdots\alpha_{j-1/2}}_{j,P,F,j_l,s_l,\rho/\lambda}(x)$ couples.

We can also use the method of QCD sum rule~\cite{Dai:1993kt,Dai:1996yw,Dai:1996qx,Dai:2003yg} to calculate Eq.~(\ref{eq:pi}) at the quark and gluon level.
Here we use $J_{1/2,-,\Xi_c,1,1,\rho}$
as an example. We first insert Eq.~(\ref{eq:current15}) into Eq.~(\ref{eq:pi}), then perform the Borel transformation, and finally obtain
\begin{eqnarray}
&& \Pi_{1/2,-,\Xi_c,1,1,\rho}(\omega_c, T) = f_{\Xi_c,1,1,\rho}^2 e^{-2 \overline{\Lambda}_{\Xi_c,1,1,\rho} / T}
\label{eq:ope}
\\ \quad && =\int_{2m_s}^{\omega_c} [ \frac{3}{4480\pi^4}\omega^7-  \frac{3m_s^2}{128\pi^4}\omega^5]e^{-\omega/T}d\omega  - \frac{3m_s\langle \bar q q \rangle}{4 \pi^2} T^4 +\frac{3m_s\langle \bar s s \rangle}{2 \pi^2} T^4- \frac{\langle g_s^2 GG \rangle}{64 \pi^4} T^4 +\frac{3m_s^2\langle g_s^2 GG \rangle}{256 \pi^4} T^2
\nonumber\\ \quad &&
\nonumber
\quad+ \frac{\langle g_s \bar q \sigma Gq \rangle\langle \bar s s \rangle}{4}+ \frac{\langle g_s \bar s \sigma Gs \rangle\langle \bar q q \rangle}{4} -\frac{m_s\langle \bar s s \rangle\langle g_s^2 GG \rangle}{128 \pi^2}-\frac{\langle g_s \bar q \sigma Gq \rangle\langle g_s \bar s \sigma Gs \rangle}{8}{1 \over T^2} \, .
\end{eqnarray}
We note that in the calculations the
software {\it Mathematica} with a package called
$FeynCalc$ is used~\cite{feyncalc}. Sum rules for other currents with different quark contents are shown in Appendix.~\ref{sec:others}.
In these equations the radiative corrections as well as the difference between up and down quarks are not taken into account
in order to simplify our calculations.
We also note that we put the condensates out of the integration to be consistent with Ref.~\cite{Dai:1993kt,Dai:1996yw,Dai:1996qx,Dai:2003yg}.
We can also put them inside the integration, but the obtained results are just slightly different from the current results.
The condensates and other parameters contained in these sum rules take the following
values~\cite{Dai:1993kt,Dai:1996yw,Dai:1996qx,Dai:2003yg,pdg,Yang:1993bp,Hwang:1994vp,Narison:2002pw,Gimenez:2005nt,Jamin:2002ev,Ioffe:2002be,Ovchinnikov:1988gk,colangelo}:
%
\begin{eqnarray}
\nonumber && \langle \bar qq \rangle = \langle \bar uu \rangle = \langle \bar dd \rangle = - (0.24 \mbox{ GeV})^3 \, ,
\\ \nonumber && \langle \bar ss \rangle = (0.8\pm 0.1)\times \langle\bar qq \rangle \, ,
\\ \nonumber &&\langle g_s^2GG\rangle =(0.48\pm 0.14) \mbox{ GeV}^4\, ,
\\ \label{condensates} && m_s = 0.15 \mbox{ GeV} \, ,
\\
\nonumber && \langle g_s \bar q \sigma G q \rangle = M_0^2 \times \langle \bar qq \rangle\, ,
\\
\nonumber && \langle g_s \bar s \sigma G s \rangle = M_0^2 \times \langle \bar ss \rangle\, ,
\\
\nonumber && M_0^2= 0.8 \mbox{ GeV}^2\, .
\end{eqnarray}

Finally, we differentiate Log[Eq.~(\ref{eq:ope})] with respect to $[-2/T]$ to obtain $\overline{\Lambda}_{j,P,F,j_l,s_l,\rho/\lambda}$,
\begin{equation}
\overline{\Lambda}_{j,P,F,j_l,s_l,\rho/\lambda}(\omega_c, T) = \frac{\frac{\partial}{\partial(-2/T)}\Pi_{j,P,F,j_l,s_l,\rho/\lambda}(\omega_c, T)}{\Pi_{j,P,F,j_l,s_l,\rho/\lambda}(\omega_c, T)} \, ,
\label{eq:mass}
\end{equation}
and use it to further evaluate $f_{j,P,F,j_l,s_l,\rho/\lambda}$:
\begin{equation}
f_{j,P,F,j_l,s_l,\rho/\lambda}(\omega_c, T) = \sqrt{\Pi_{j,P,F,j_l,s_l,\rho/\lambda}(\omega_c, T) \times e^{2 \overline{\Lambda}_{j,P,F,j_l,s_l,\rho/\lambda}(\omega_c, T) / T}} \, .
\label{eq:coupling}
\end{equation}

As noted above, the two currents in the same doublet give identical sum rules at the leading order, so we have $\overline{\Lambda}_{j_l+1/2,P,F,j_l,s_l,\rho/\lambda} = \overline{\Lambda}_{|j_l-1/2|,P,F,j_l,s_l,\rho/\lambda}$. Moreover, the $P$-wave charmed baryons always have a negative parity. To slightly simplify our notations, we use another symbol $\overline{\Lambda}_{F,j_l,s_l,\rho/\lambda}$ to denote them. Similarly, we use the symbol $m_{F,j_l,s_l,\rho/\lambda}$ to denote  $m_{j_l+1/2,P,F,j_l,s_l,\rho/\lambda}$ and $m_{|j_l-1/2|,P,F,j_l,s_l,\rho/\lambda}$, and $f_{F,j_l,s_l,\rho/\lambda}$ to denote $f_{j_l+1/2,P,F,j_l,s_l,\rho/\lambda}$ and $f_{|j_l-1/2|,P,F,j_l,s_l,\rho/\lambda}$. This is also ``true'' for those currents contained in baryon singlets. We note that we work at the $O(1/m_Q)$ order in the next section to differentiate the two currents within the same doublet.

Then we start to perform the numerical analysis, and we have altogether three criteria. The first criterion is to require that the high-order power corrections be less than 30\%:
%
\begin{equation}
\label{eq_convergence}
\mbox{Convergence (CVG)} \equiv |\frac{ \Pi^{\rm high-order}_{j,P,F,j_l,s_l,\rho/\lambda}(\omega_c, T) }{ \Pi_{j,P,F,j_l,s_l,\rho/\lambda}(\omega_c, T) }| \leq 30\% \, ,
\end{equation}
%
where $\Pi^{\rm high-order}_{j,P,F,j_l,s_l,\rho/\lambda}(\omega_c, T)$ is used to denote the high-order power corrections, for example,
%
\begin{eqnarray}
\Pi^{\rm high-order}_{1/2,-,\Xi_c,1,1,\rho}(\omega_c, T) &=& - \frac{3m_s\langle \bar q q \rangle}{4 \pi^2} T^4 +\frac{3m_s\langle \bar s s \rangle}{2 \pi^2} T^4- \frac{\langle g_s^2 GG \rangle}{64 \pi^4} T^4 +\frac{3m_s^2\langle g_s^2 GG \rangle}{256 \pi^4} T^2
\nonumber \\ \quad &&
+ \frac{\langle g_s \bar q \sigma Gq \rangle\langle \bar s s \rangle}{4}+ \frac{\langle g_s \bar s \sigma Gs \rangle\langle \bar q q \rangle}{4} -\frac{m_s\langle \bar s s \rangle\langle g_s^2 GG \rangle}{128 \pi^2}-\frac{\langle g_s \bar q \sigma Gq \rangle\langle g_s \bar s \sigma Gs \rangle}{8}{1 \over T^2} \, .
\end{eqnarray}
%
The second criterion is to require that the pole contribution (PC) be larger than 20\%:
%
\begin{equation}
\label{eq_pole}
\mbox{PC} \equiv \frac{ \Pi_{j,P,F,j_l,s_l,\rho/\lambda}(\omega_c, T) }{ \Pi_{j,P,F,j_l,s_l,\rho/\lambda}( \infty , T) } \geq 20\% \, .
\end{equation}
%
Then we obtain an interval $T_{min}<T<T_{max}$ for a fixed $\omega_c$. This $\omega_c$ is a free parameter, which is chosen to be around 3.6 GeV in order to fit the masses of $\Xi_c(2790)$ ($J^P=1/2^-$) and $\Xi_c(2815)$ ($J^P=3/2^-$)~\cite{pdg}. In this case an interval $0.45$ GeV $< T < 0.64$ GeV is obtained for $\omega_c = 3.6$ GeV.

\begin{figure}[hbt]
\begin{center}
\scalebox{0.6}{\includegraphics{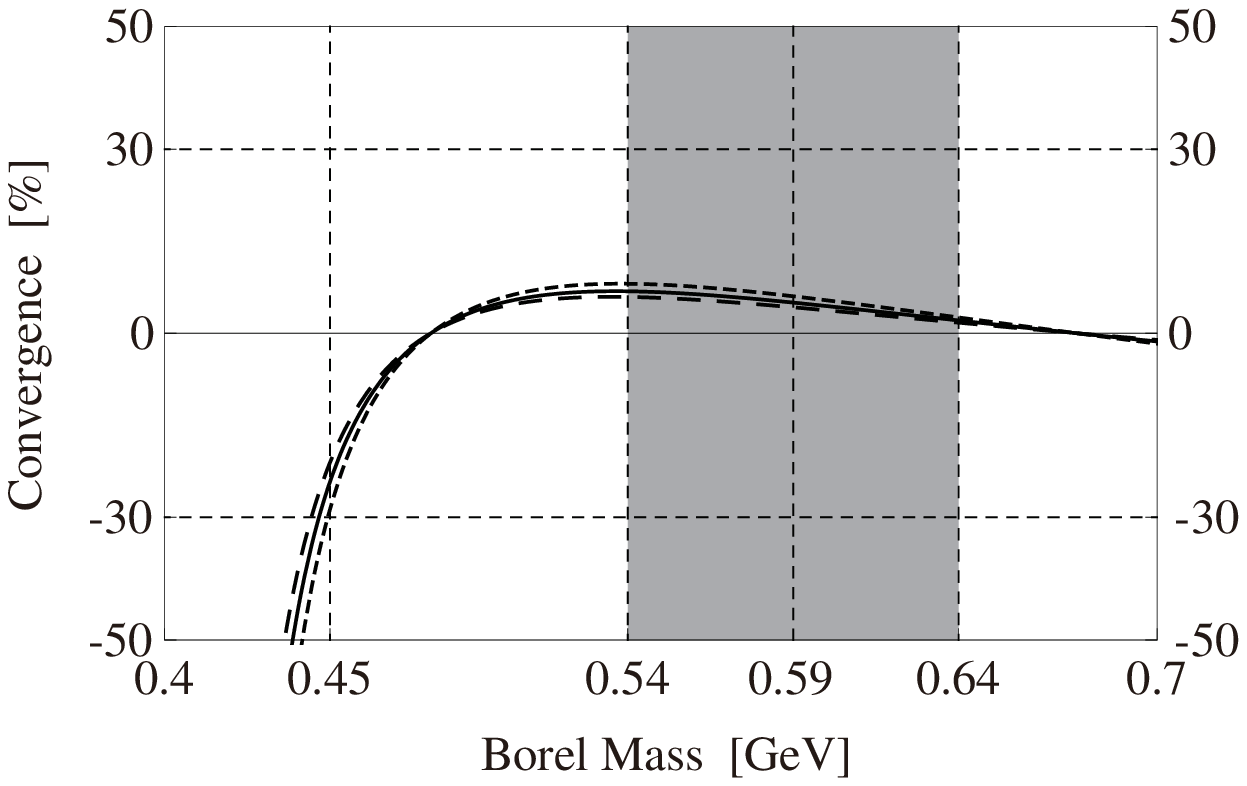}}
\scalebox{0.573}{\includegraphics{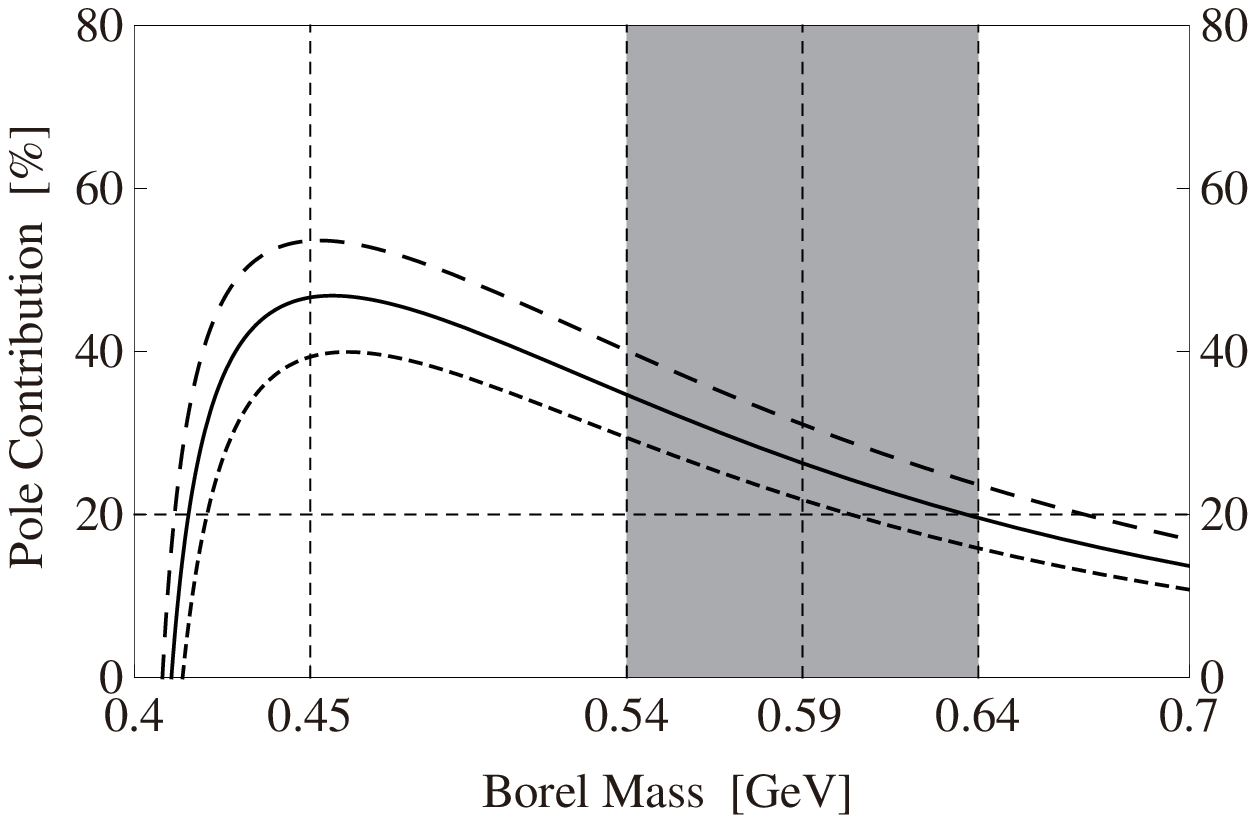}}
\caption{The variations of CVG and PC, as defined in Eqs.~(\ref{eq_convergence}) and (\ref{eq_pole}), with respect to the Borel mass $T$, when $J_{1/2,-,\Xi_c,1,1,\rho}$ is used. We obtain the short-dashed, solid, and long-dashed curves by fixing $\omega_c = 3.4$, 3.6, and 3.8 GeV, respectively.}
\label{fig:pole}
\end{center}
\end{figure}

\begin{figure}[hbt]
\begin{center}
\scalebox{0.597}{\includegraphics{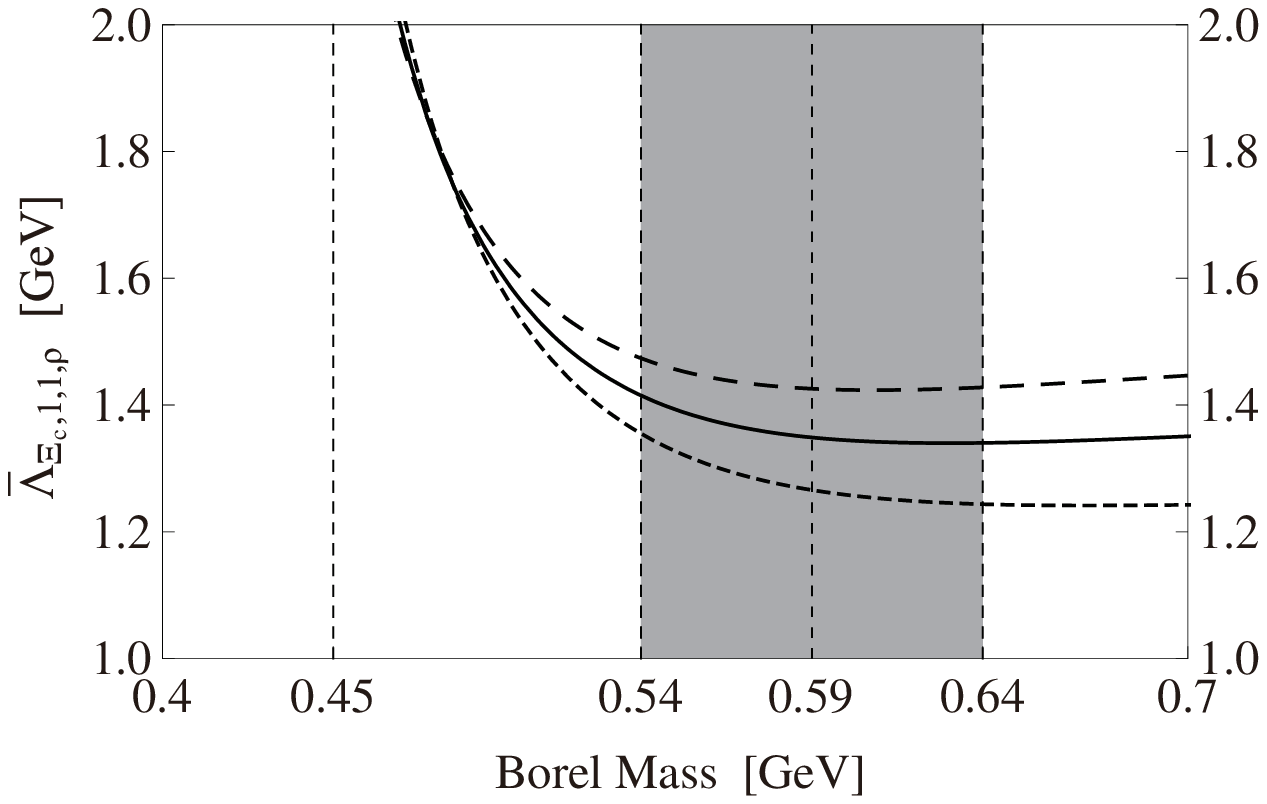}}
\scalebox{0.6}{\includegraphics{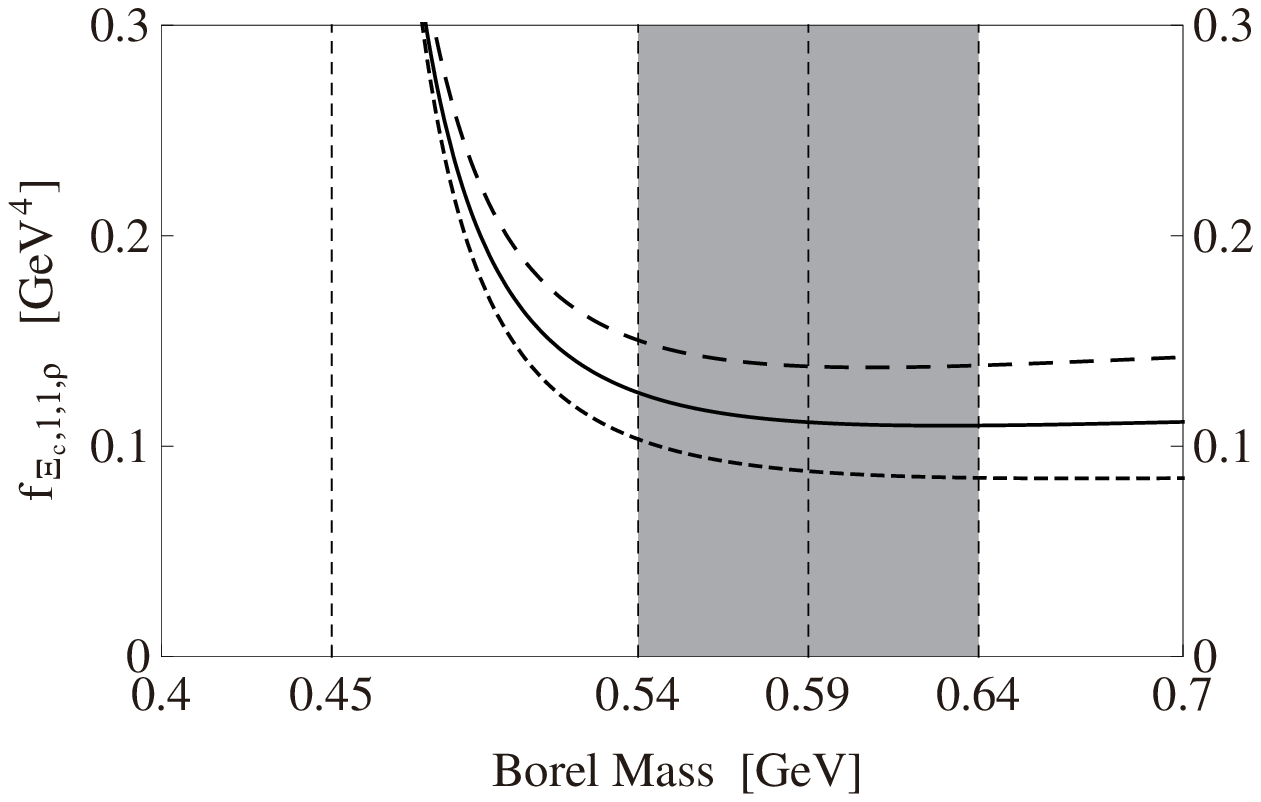}} \caption{The
variations of $\overline{\Lambda}_{\Xi_c,1,1,\rho}$ (left) and $f_{\Xi_c,1,1,\rho}$ (right) with respect to the
Borel mass $T$, when $J_{1/2,-,\Xi_c,1,1,\rho}$ is used. We use the working region $0.54$ GeV $< T < 0.64$ GeV, and obtain the short-dashed, solid, and long-dashed curves by fixing $\omega_c = 3.4$, 3.6, and 3.8 GeV, respectively.} \label{fig:leading}
\end{center}
\end{figure}

To see this clearly, we show the variations of CVG and PC, as defined in Eqs.~(\ref{eq_convergence}) and (\ref{eq_pole}), with respect to the Borel mass $T$ in Fig.~\ref{fig:pole}, and the variations of $\overline{\Lambda}_{\Xi_c,1,1,\rho}$ and $f_{\Xi_c,1,1,\rho}$ also with respect to $T$ in Fig.~\ref{fig:leading}. From these figures, we find that as the curve of CVG quickly increases to its top point around $T = 0.54$ GeV, the dependence of $\overline{\Lambda}_{\Xi_c,1,1,\rho}$ and $f_{\Xi_c,1,1,\rho}$ on the Borel mass $T$ is significant; while as this curve slowly decreases from the top point, the dependence becomes much weaker. Accordingly, our third criterion is to require that this dependence be weak (see Figs.~\ref{fig:Xi10} and \ref{fig:Xi21} for more examples). Finally, we use the new interval $0.54$ GeV $< T < 0.64$ GeV as our working region, during which the following numerical results are obtained:
\begin{eqnarray}
\overline{\Lambda}_{\Xi_c,1,1,\rho} &=& 1.35 \pm 0.13 \mbox{ GeV} \, ,
\\ f_{\Xi_c,1,1,\rho} &=& 0.11 \pm 0.04 \mbox{ GeV}^{4} \, .
\end{eqnarray}
In these equations the central values are obtained by fixing $T=0.59$ GeV and $\omega_c = 3.6$ GeV.

\section{The Sum Rules at the ${\mathcal O}(1/m_Q)$ Order}
\label{sec:nexttoleading}

In this section we work up to the ${\mathcal O}(1/m_Q)$ order. To do this, we use the following Lagrangian of HQET~\cite{Dai:1996qx,Dai:2003yg}:
\begin{eqnarray}
\mathcal{L}_{\rm eff} = \overline{h}_{v}iv\cdot Dh_{v} + \frac{1}{2m_{Q}}\mathcal{K} + \frac{1}{2m_{Q}}\mathcal{S} \, .
\label{eq:next}
\end{eqnarray}
In this Lagrangian we use $\mathcal{K}$ to denote the operator of nonrelativistic kinetic energy,
\begin{eqnarray}
\mathcal{K} = \overline{h}_{v}(iD_{t})^{2}h_{v} \, ,
\end{eqnarray}
and use $\mathcal S$ to denote the Pauli term describing the chromomagnetic interaction:
\begin{eqnarray}
\mathcal{S}= \frac{g}{2} C_{mag} (m_{Q}/\mu) \overline{h}_{v} \sigma_{\mu\nu} G^{\mu\nu} h_{v} \, ,
\end{eqnarray}
where the two parameters are $C_{mag} (m_{Q}/\mu) = [ \alpha_s(m_Q) / \alpha_s(\mu) ]^{3/\beta_0}$ and $\beta_0 = 11 - 2 n_f /3$.

We can write the pole term up to the ${\mathcal O}(1/m_Q)$ order as
\begin{eqnarray}
\Pi(\omega)_{pole} &=& \frac{(f+\delta f)^{2}}{2(\overline{\Lambda}+\delta m)-\omega}
\label{eq:correction}
\\ \nonumber &=& \frac{f^{2}}{2\overline{\Lambda}-\omega}-\frac{2\delta mf^{2}}{(2\overline{\Lambda}-\omega)^{2}}+\frac{2f\delta f}{2\overline{\Lambda}-\omega} \, ,
\end{eqnarray}
where $\delta m_{F,j_l,s_l,\rho/\lambda}$ and $\delta f_{F,j_l,s_l,\rho/\lambda}$ are the corrections to the mass $m_{F,j_l,s_l,\rho/\lambda}$ and the coupling constant $f_{F,j_l,s_l,\rho/\lambda}$. In this paper we shall not consider $\delta f$. To calculate $\delta m$, we consider the following three-point correlation functions:
\begin{eqnarray}
\delta_{O}\Pi_{j,P,F,j_l,s_l,\rho/\lambda}^{\alpha_{1}\cdots\alpha_{j-1/2},\beta_{1}\cdots\beta_{j-1/2}}(\omega , \omega ')
&=& i^{2}\int d^{4}xd^{4}ye^{ik\cdot x-ik'\cdot y}\times\langle0|T[J_{j,P,F,j_l,s_l,\rho/\lambda}^{\alpha_{1}\cdots \alpha_{j-1/2}}(x)O(0) \bar J_{j,P,F,j_l,s_l,\rho/\lambda}^{\beta_{1}\cdots \beta_{j-1/2}}(y)]|0\rangle
\label{eq:nextpi}
\\ \nonumber &=& \mathbb{S} [ g_t^{\alpha_1 \beta_1} \cdots g_t^{\alpha_{j-1/2} \beta_{j-1/2}} ] \delta_{O} \Pi_{j,P,F,j_l,s_l,\rho/\lambda} (\omega) \, ,
\end{eqnarray}
where $O = \mathcal{K}$ or $\mathcal{S}$, and $\mathbb{S}[\cdots]$ is used to denote symmetrization and subtracting the trace
terms in the sets $(\alpha_1 \cdots \alpha_{j-1/2})$ and $(\beta_1 \cdots
\beta_{j-1/2})$. Based on the Lagrangian (\ref{eq:next}), we can write Eqs.~(\ref{eq:nextpi}) at the hadron level as
\begin{eqnarray}
\delta_{\mathcal{K}}\Pi(\omega,\omega')_{j,P,F,j_l,s_l,\rho/\lambda} &=& \frac{f^{2}K_{F,j_l,s_l,\rho/\lambda}}{(2\overline{\Lambda}-\omega)(2\overline{\Lambda}-\omega')}
+\frac{f^{2}G_{\mathcal{K}}(\omega')}{2\overline{\Lambda}-\omega}
+\frac{f^{2}G_{\mathcal{K}}(\omega)}{2\overline{\Lambda}-\omega'} \, ,
\label{eq:K}
\\ \delta_{\mathcal{S}}\Pi(\omega,\omega')_{j,P,F,j_l,s_l,\rho/\lambda} &=& \frac{d_{M}f^{2}\Sigma_{F,j_l,s_l,\rho/\lambda}}{(2\overline{\Lambda}-\omega)(2\overline{\Lambda}-\omega')}
+\frac{d_{M}f^{2}G_{\mathcal{S}}(\omega')}{2\overline{\Lambda}-\omega} \,
+\frac{d_{M}f^{2}G_{\mathcal{S}}(\omega)}{2\overline{\Lambda}-\omega'} \, ,
\label{eq:S}
\end{eqnarray}
where $K_{F,j_l,s_l,\rho/\lambda}$, $\Sigma_{F,j_l,s_l,\rho/\lambda}$ and $d_{M}$ are defined to be
\begin{eqnarray}
\nonumber K_{F,j_l,s_l,\rho/\lambda} &\equiv& \langle j,P,F,j_l,s_l,\rho/\lambda|\overline{h}_{v}(iD_{\bot})^{2}h_{v}|j,P,F,j_l,s_l,\rho/\lambda\rangle \, ,
\\ \nonumber d_{M}\Sigma_{F,j_l,s_l,\rho/\lambda} &\equiv& \langle j,P,F,j_l,s_l,\rho/\lambda| {g\over2} \overline{h}_{v}\sigma_{\mu\nu}G^{\mu\nu}h_{v}|j,P,F,j_l,s_l,\rho/\lambda\rangle \, ,
\\ d_{M} &\equiv& d_{j,j_{l}} \, ,
\\ \nonumber d_{j_{l}-1/2,j_{l}} &=& 2j_{l}+2\, ,
\\ \nonumber d_{j_{l}+1/2,j_{l}} &=& -2j_{l} \, .
\end{eqnarray}
We note that the term $\mathcal S$ can cause a mass splitting within the same doublet as well as a mixing of states with the same $j$, $P$, $F$ but different $j_l$, $s_l$, $\rho/\lambda$. For example, $|3/2,-,\mathbf{\bar3}_F,1,1,\rho\rangle$ and $|3/2,-,\mathbf{\bar3}_F,2,1,\rho\rangle$ can be mixed due to this term. However, this effect is found to be negligible in Ref.~\cite{Dai:1998ve}, so we do not consider this effect in this paper.
Then we fix $\omega = \omega^\prime$ and use Eqs.~(\ref{eq:correction}), (\ref{eq:K}), and (\ref{eq:S}) to obtain
\begin{eqnarray}
\delta m_{F,j_l,s_l,\rho/\lambda} = -\frac{1}{4m_{Q}}(K_{F,j_l,s_l,\rho/\lambda} + d_{M}C_{mag}\Sigma_{F,j_l,s_l,\rho/\lambda} ) \, .
\end{eqnarray}

\begin{figure}[hbt]
\begin{center}
\scalebox{0.6}{\includegraphics{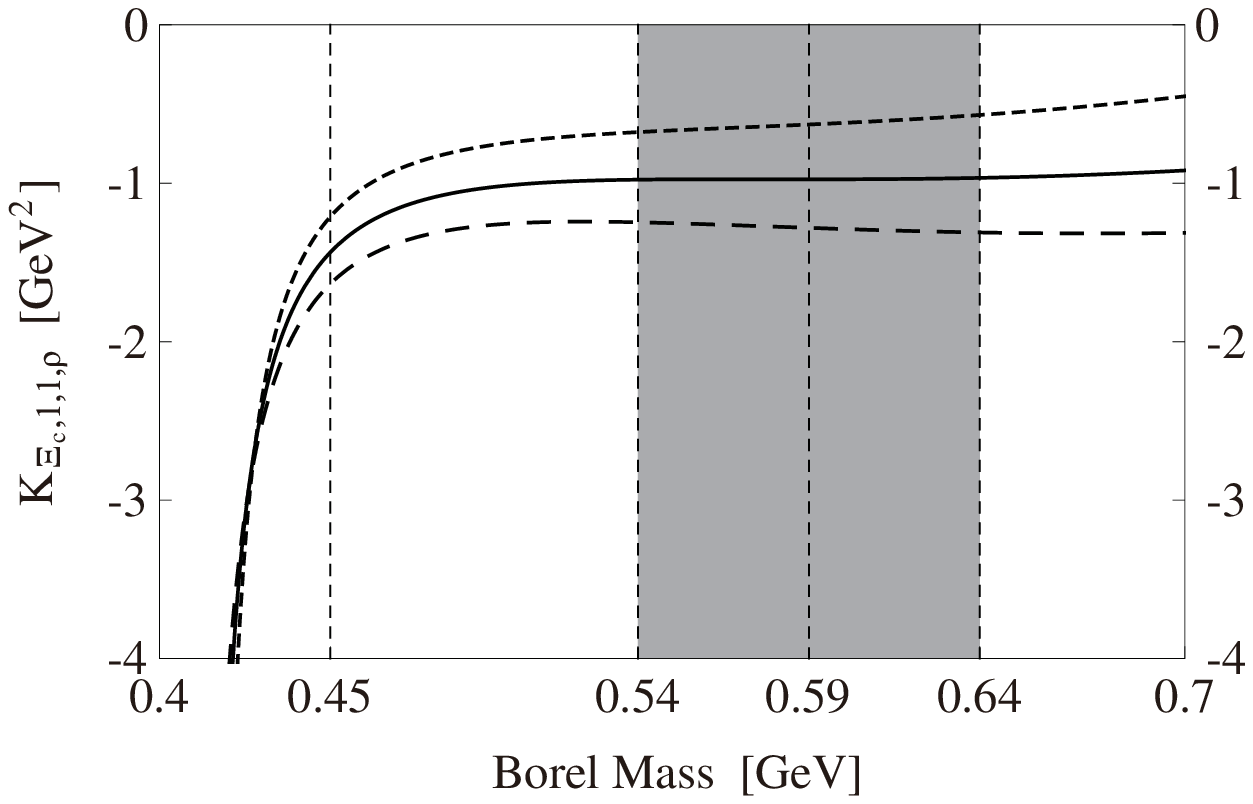}}
\scalebox{0.592}{\includegraphics{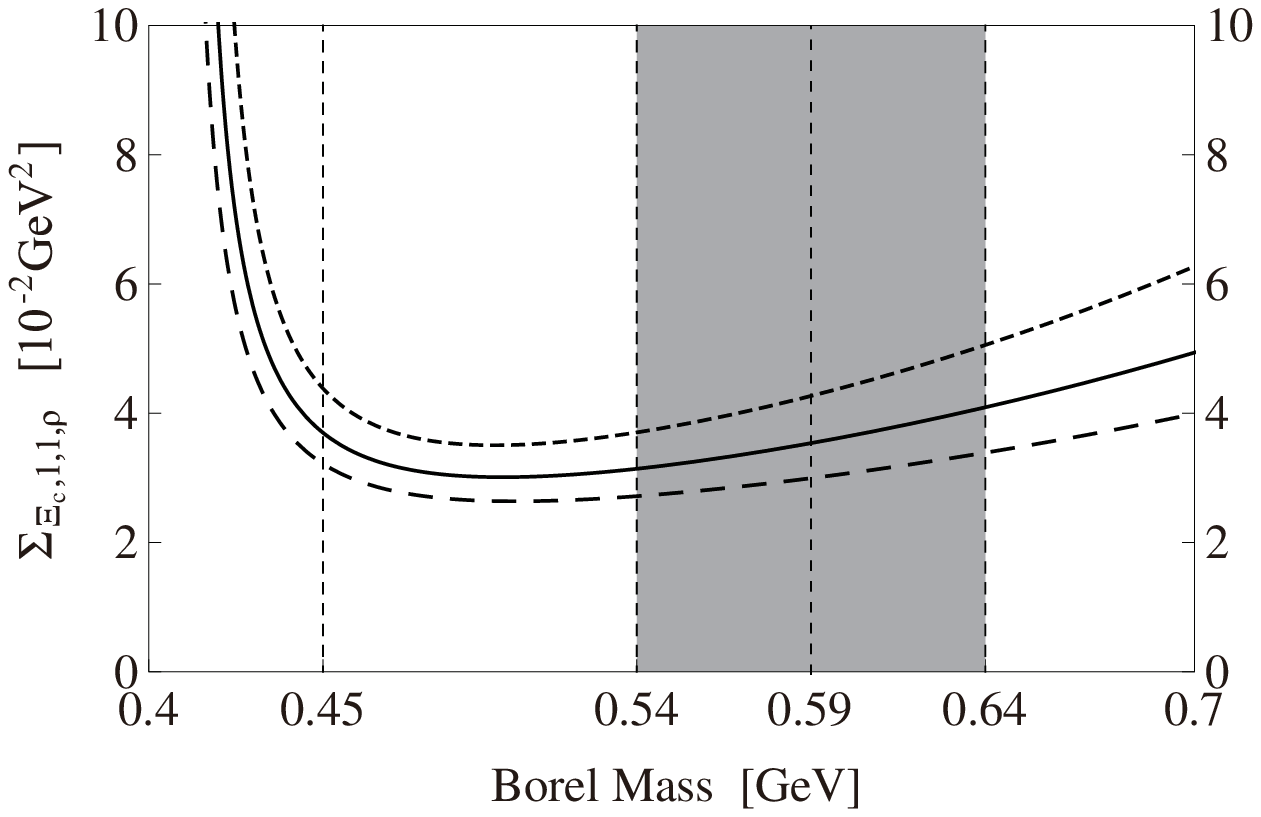}} \caption{The variations
of $K_{\Xi_c,1,1,\rho}$ (left) and $\Sigma_{\Xi_c,1,1,\rho}$ (right) with respect to the Borel mass
$T$, when $J_{1/2,-,\Xi_c,1,1,\rho}$ is used. We use the working region $0.54$ GeV $< T < 0.64$
GeV, and obtain the short-dashed, solid and
long-dashed curves by fixing $\omega_c = 3.4$, 3.6 and
3.8 GeV, respectively.} \label{fig:next}
\end{center}
\end{figure}

We can also calculate Eqs.~(\ref{eq:nextpi}) at the quark and gluon level using the method of QCD sum rule~\cite{Dai:1996qx,Dai:2003yg}.
Again we use $J_{1/2,-,\Xi_c,1,1,\rho}$ as an example.
After inserting Eq.~(\ref{eq:current15}) into Eqs.~(\ref{eq:nextpi}), and making a double Borel transformation for both $\omega$ and $\omega^\prime$,
we obtain sum rules having two Borel parameters $T_1$ and $T_2$; then taking these two Borel parameters to be
equal, we obtain the following two sum rules for $K_{\Xi_c,1,1,\rho}$ and
$\Sigma_{\Xi_c,1,1,\rho}$:
\begin{eqnarray}
&& f_{\Xi_c,1,1,\rho}^2 K_{\Xi_c,1,1,\rho} e^{-2 \overline{\Lambda}_{\Xi_c,1,1,\rho} / T}
\label{eq:Kc}
\\ \quad && =\int_{2m_s}^{\omega_c} [ -\frac{1}{6720\pi^4}\omega^9+  \frac{13m_s^2}{1792\pi^4}\omega^7]e^{-\omega/T}d\omega + \frac{15m_s\langle \bar q q \rangle}{2 \pi^2} T^6- \frac{33m_s\langle \bar s s \rangle}{2 \pi^2} T^6+\frac{9\langle g_s^2 GG \rangle}{64 \pi^4} T^6-\frac{3m_s\langle g_s \bar q \sigma Gq \rangle}{2 \pi^2} T^4
\nonumber\\ \quad &&\quad
-\frac{21m_s^2\langle g_s^2 GG \rangle}{256 \pi^4} T^4-\frac{\langle g_s \bar q \sigma G q\rangle\langle g_s \bar s \sigma Gs \rangle}{8}-\frac{3m_s\langle \bar q q \rangle\langle g_s^2 GG \rangle}{64 \pi^2} T^2+\frac{m_s\langle \bar s s \rangle\langle g_s^2 GG \rangle}{32\pi^2} T^2+\frac{\langle \bar q q \rangle\langle g_s \bar s \sigma Gs \rangle\langle g_s^2 GG \rangle}{64}{1 \over T^2}
\nonumber\\ \quad &&\quad
+\frac{\langle \bar s s \rangle\langle g_s \bar q \sigma Gq \rangle\langle g_s^2 GG \rangle}{64}{1 \over T^2}-\frac{\langle g_s \bar q \sigma Gq \rangle\langle g_s \bar s \sigma Gs \rangle\langle g_s^2 GG \rangle}{128}{1 \over T^4}\, ,
\nonumber \\
&& f_{\Xi_c,1,1,\rho}^2 \Sigma_{\Xi_c,1,1,\rho} e^{-2 \overline{\Lambda}_{\Xi_c,1,1,\rho} / T}
=\frac{3\langle g_s^2 GG \rangle}{32 \pi^4} T^6- \frac{5m_s^2\langle g_s^2 GG \rangle}{128 \pi^4} T^4+\frac{m_s\langle \bar s s \rangle\langle g_s^2 GG \rangle}{48 \pi^2} T^2\, .
\label{eq:Sc}
\end{eqnarray}
Sum rules for other currents with different quark contents are shown in Appendix~\ref{sec:others}.
To obtain $K_{\Xi_c,1,1,\rho}$ and
$\Sigma_{\Xi_c,1,1,\rho}$, we just need to simply divide Eqs.~(\ref{eq:Kc}) and (\ref{eq:Sc}) by the sum rule (\ref{eq:ope}). Their variations
are shown in Fig.~\ref{fig:next} with respect to the Borel mass $T$, and their dependence on $T$ is found to be weak in our working region $0.54$ GeV $< T < 0.64$ GeV, during which we obtain the following numerical results:
\begin{eqnarray}
K_{\Xi_c,1,1,\rho} &=& -0.98 \pm 0.41 \mbox{ GeV}^2 \, ,
\\ \Sigma_{\Xi_c,1,1,\rho} &=& 0.035 \pm 0.015 \mbox{ GeV}^{2} \, .
\end{eqnarray}
In these equations the central values are obtained by fixing $T=0.59$ GeV and $\omega_c = 3.6$ GeV. We note that this dependence
is also weak in the interval $0.45$ GeV $< T < 0.64$ GeV demanded by the first two criteria, but the numerical results
obtained in this interval are almost the same as those obtained in the working region $0.54$ GeV $< T < 0.64$ GeV.

\section{Numerical Results and Discussions}
\label{sec:summary}

We can obtain the weighted
average mass for the heavy baryon doublet $[\Xi_c,1,1,\rho]$ using those results obtained in Secs.~\ref{sec:leading}-\ref{sec:nexttoleading}:
\begin{eqnarray}
{1\over6} \Big ( 2 m_{\Xi_c({1/2}^-)} + 4 m_{\Xi_c({3/2}^-)} \Big) &=& m_c + (1.35 \pm 0.13) \mbox{ GeV } - {1 \over 4m_c} [ (-0.98 \pm 0.41)
\mbox{ GeV}^2 ] \, ,
\end{eqnarray}
where $\Xi_c({1/2}^-)$ and $\Xi_c({3/2}^-)$ are the two baryons contained in this doublet.
At the same time we can also obtain their mass splitting:
\begin{eqnarray}
m_{\Xi_c({3/2}^-)} - m_{\Xi_c({1/2}^-)} &=& {1 \over 4m_c} \times {6} \times [ (0.035 \pm 0.015) \mbox{ GeV}^2 ] \, .
\end{eqnarray}
Clearly we find that we should not neglect the $\mathcal{O}(1/m_Q)$ corrections.

Then we use the PDG value $m_c = 1.275 \pm 0.025$ GeV~\cite{pdg} for the charm quark mass in the $\overline{\rm MS}$ scheme
to obtain numerical results. Its pole mass may also be used, but then we need to properly fine-tune the threshold value $\omega_c$
to fit the physical mass values. This suggests that there are large theoretical uncertainties in our results for the masses of the heavy baryons. However,
their differences within the same doublet are produced quite well with much less theoretical uncertainty because they do not depend much on the charm quark mass and the threshold value:
\begin{eqnarray}
\nonumber m_{\Xi_c({1/2}^-)} &=& 2.79 \pm 0.15 \mbox{ GeV} \, ,
\\ m_{\Xi_c({3/2}^-)} &=& 2.83 \pm 0.15 \mbox{ GeV} \, ,
\\ \nonumber m_{\Xi_c({3/2}^-)} - m_{\Xi_c({1/2}^-)} &=& 42 \pm 18 \mbox{ MeV} \, .
\end{eqnarray}
These results are consistent with the masses of $\Xi_c(2790)$ ($J^P=1/2^-$) and $\Xi_c(2815)$ ($J^P=3/2^-$) as well as their difference~\cite{pdg}:
\begin{eqnarray}
\nonumber && m^{\rm exp}_{\Xi_c(2790)^+} = 2789.1 \pm 3.2 \mbox{ MeV} \, , m^{\rm exp}_{\Xi_c(2790)^0} = 2791.8 \pm 3.3 \mbox{ MeV} \, ,
\\ && m^{\rm exp}_{\Xi_c(2815)^+} = 2816.6 \pm 0.9 \mbox{ MeV} \, , m^{\rm exp}_{\Xi_c(2815)^0} = 2819.6 \pm 1.2 \mbox{ MeV} \, ,
\\ \nonumber && m^{\rm exp}_{\Xi_c(2815)} - m^{\rm exp}_{\Xi_c(2790)} \approx 28 \mbox{ MeV} \, .
\end{eqnarray}
Besides these two states, there are five other well-observed states, which may be $P$-wave charm baryons. They are $\Lambda_c(2595)$ ($J^P=1/2^-$), $\Lambda_c(2625)$ ($J^P=3/2^-$), $\Sigma_c(2800)$ ($J^P=?^?$), $\Xi_c(2980)$ ($J^P=?^?$) and $\Xi_c(3080)$ ($J^P=?^?$)~\cite{pdg}. Their masses are
\begin{eqnarray}
\nonumber && m^{\rm exp}_{\Lambda_c(2595)^+} = 2592.25 \pm 0.28 \mbox{ MeV} \, , m^{\rm exp}_{\Lambda_c(2625)^+} = 2628.11 \pm 0.19 \mbox{ MeV} \, ,
\\ \nonumber && m^{\rm exp}_{\Lambda_c(2625)^+} - m^{\rm exp}_{\Lambda_c(2595)^+} \approx 36 \mbox{ MeV} \, ,
\\ \nonumber && m^{\rm exp}_{\Sigma_c(2800)^{++}} = 2801^{+4}_{-6} \mbox{ MeV} \, , m^{\rm exp}_{\Sigma_c(2800)^{+}} = 2792^{+14}_{-5} \mbox{ MeV} \, , m^{\rm exp}_{\Sigma_c(2800)^{0}} = 2806^{+5}_{-7} \mbox{ MeV} \, ,
\\ \nonumber && m^{\rm exp}_{\Xi_c(2980)^+} = 2971.4 \pm 3.3 \mbox{ MeV} \, , m^{\rm exp}_{\Xi_c(2980)^0} = 2968.0 \pm 2.6 \mbox{ MeV} \, ,
\\ \nonumber && m^{\rm exp}_{\Xi_c(3080)^+} = 3077.0 \pm 0.4 \mbox{ MeV} \, , m^{\rm exp}_{\Xi_c(3080)^0} = 3079.9 \pm 1.4 \mbox{ MeV} \,
\\ \nonumber && m^{\rm exp}_{\Xi_c(3080)} - m^{\rm exp}_{\Xi_c(2980)} \approx 109 \mbox{ MeV} \, .
\end{eqnarray}
We use the baryon multiplets $[\mathbf{\bar 3}_F, 0/1, 0/1, \rho/\lambda]$ to fit the states $\Lambda_c(2595)$ ($J^P=1/2^-$), $\Lambda_c(2625)$ ($J^P=3/2^-$), $\Xi_c(2790)$ ($J^P=1/2^-$) and $\Xi_c(2815)$ ($J^P=3/2^-$), and use the multiplets $[\mathbf{6}_F, 0/1, 0/1, \rho/\lambda]$ to fit the states $\Sigma_c(2800)$ ($J^P=?^?$), $\Xi_c(2980)$ ($J^P=?^?$) and $\Xi_c(3080)$ ($J^P=?^?$). The procedures are just the same as before. We do not discuss the details any more, but summarize the good fitting results in Table~\ref{tab:results1} and other results in Table~\ref{tab:results2}. Considering there are no excited $\Omega_c$ observed in experiments, we assume that free parameters $\omega_c$ in the same multiplet satisfy the relation $\omega_c(\Omega_c) - \omega_c(\Xi^\prime_c) = \omega_c(\Xi^\prime_c)- \omega_c(\Sigma_c)$, and use $\omega_c(\Omega_c)$ to evaluate masses of $\Omega_c$ baryons. We note that this difference is 0.5 GeV for the three baryon multiplets $[\mathbf{6}_F, 1, 0, \rho]$, $[\mathbf{6}_F, 2, 1, \lambda]$, and $[\mathbf{6}_F, 0, 1, \lambda]$ among four $[\mathbf{6}_F, 0/1, 0/1, \rho/\lambda]$ multiplets.

\begin{table}[hbt]
\begin{center}
\caption{We use the baryon multiplets $[\mathbf{\bar 3}_F, 1, 1, \rho]$ to fit the states $\Lambda_c(2595)$ ($J^P=1/2^-$), $\Lambda_c(2625)$ ($J^P=3/2^-$), $\Xi_c(2790)$ ($J^P=1/2^-$) and $\Xi_c(2815)$ ($J^P=3/2^-$), and use the multiplets $[\mathbf{6}_F, 1, 0, \rho]$ and $[\mathbf{6}_F, 2, 1, \lambda]$ to fit the states $\Sigma_c(2800)$ ($J^P=?^?$), $\Xi_c(2980)$ ($J^P=?^?$) and $\Xi_c(3080)$ ($J^P=?^?$). The procedures are the same as before, and the good fitting results are summarized here. We assume that free parameters $\omega_c$ in the same multiplet satisfy the relation $\omega_c(\Omega_c) - \omega_c(\Xi^\prime_c) = \omega_c(\Xi^\prime_c)- \omega_c(\Sigma_c)$, and use $\omega_c(\Omega_c)$ to evaluate the mass of $\Omega_c$.}
\begin{tabular}{c | c | c | c | c c | c c | c c | c}
\hline\hline
\multirow{2}{*}{Multiplets} & \multirow{2}{*}{B} & $\omega_c$ & Working region & $\overline{\Lambda}$ & $f$ & $K$ & $\Sigma$ & Baryons & Mass & Difference
\\ & & (GeV) & (GeV) & (GeV) & (GeV$^{4}$) & (GeV$^2$) & (GeV$^2$) & ($j^P$) & (GeV) & (MeV)
\\ \hline\hline
\multirow{4}{*}{$[\mathbf{\bar 3}_F, 1, 1, \rho]$} & \multirow{2}{*}{$\Lambda_c$} & \multirow{2}{*}{3.1} & \multirow{2}{*}{$0.54< T < 0.59$} & \multirow{2}{*}{$1.16 \pm 0.13$} & \multirow{2}{*}{$0.07 \pm 0.03$} & \multirow{2}{*}{$-0.99 \pm 0.24$} & \multirow{2}{*}{$0.042\pm0.014$}
& $\Lambda_c(1/2^-)$ & $2.60 \pm 0.14$ & \multirow{2}{*}{$49 \pm 16$}
\\ \cline{9-10}
& & & & & & & & $\Lambda_c(3/2^-)$ & $2.65 \pm 0.14$ &
\\ \cline{2-11}
& \multirow{2}{*}{$\Xi_c$} & \multirow{2}{*}{3.6} & \multirow{2}{*}{$0.54< T < 0.64$} & \multirow{2}{*}{$1.35 \pm 0.13$} & \multirow{2}{*}{$0.11 \pm 0.04$} & \multirow{2}{*}{$-0.98 \pm 0.41$} & \multirow{2}{*}{$0.035 \pm 0.015$}
& $\Xi_c(1/2^-)$ & $2.79 \pm 0.15$ & \multirow{2}{*}{$42 \pm 18$}
\\ \cline{9-10}
& & & & & & & & $\Xi_c(3/2^-)$ & $2.83 \pm 0.15$ &
\\ \hline
\multirow{6}{*}{$[\mathbf{6}_F, 1, 0, \rho]$} & \multirow{2}{*}{$\Sigma_c$} & \multirow{2}{*}{3.4} & \multirow{2}{*}{$0.53< T < 0.64$} & \multirow{2}{*}{$1.22 \pm 0.17$} & \multirow{2}{*}{$0.06 \pm 0.03$} & \multirow{2}{*}{$-1.24 \pm 0.23$} & \multirow{2}{*}{$0.013 \pm 0.006$} & $\Sigma_c(1/2^-)$ & $2.73 \pm 0.18$ & \multirow{2}{*}{$15 \pm 7$}
\\ \cline{9-10}
& & & & & & & & $\Lambda_c(3/2^-)$ & $2.75 \pm 0.18$ &
\\ \cline{2-11}
& \multirow{2}{*}{$\Xi^\prime_c$} & \multirow{2}{*}{3.9} & \multirow{2}{*}{$0.52< T < 0.70$} & \multirow{2}{*}{$1.42 \pm 0.13$} & \multirow{2}{*}{$0.10 \pm 0.03$} & \multirow{2}{*}{$-1.40 \pm 0.37$} & \multirow{2}{*}{$0.010 \pm 0.006$}
& $\Xi^\prime_c(1/2^-)$ & $2.96 \pm 0.15$ & \multirow{2}{*}{$12 \pm 7$}
\\ \cline{9-10}
& & & & & & & & $\Xi^\prime_c(3/2^-)$ & $2.98 \pm 0.15$ &
\\ \cline{2-11}
& \multirow{2}{*}{$\Omega_c$} & \multirow{2}{*}{4.4} & \multirow{2}{*}{$0.51< T < 0.77$} & \multirow{2}{*}{$1.64 \pm 0.16$} & \multirow{2}{*}{$0.15 \pm 0.05$} & \multirow{2}{*}{$-1.71 \pm 0.57$} & \multirow{2}{*}{$0.008 \pm 0.005$}
& $\Omega_c(1/2^-)$ & $3.25 \pm 0.20$ & \multirow{2}{*}{$10 \pm 6$}
\\ \cline{9-10}
& & & & & & & & $\Omega_c(3/2^-)$ & $3.26 \pm 0.19$ &
\\ \hline
\multirow{6}{*}{$[\mathbf{6}_F, 2, 1, \lambda]$} & \multirow{2}{*}{$\Sigma_c$} & \multirow{2}{*}{3.0} & \multirow{2}{*}{$0.55< T < 0.58$} & \multirow{2}{*}{$1.10 \pm 0.13$} & \multirow{2}{*}{$0.06 \pm 0.02$} & \multirow{2}{*}{$-2.43 \pm 0.28$} & \multirow{2}{*}{$0.043 \pm 0.012$}
& $\Sigma_c(3/2^-)$ & $2.80 \pm 0.15$ & \multirow{2}{*}{$85 \pm 23$}
\\ \cline{9-10}
& & & & & & & & $\Sigma_c(5/2^-)$ & $2.89 \pm 0.15$ &
\\ \cline{2-11}
& \multirow{2}{*}{$\Xi^\prime_c$} & \multirow{2}{*}{3.5} & \multirow{2}{*}{$0.53< T < 0.64$} & \multirow{2}{*}{$1.25 \pm 0.18$} & \multirow{2}{*}{$0.08 \pm 0.04$} & \multirow{2}{*}{$-2.51 \pm 0.53$} & \multirow{2}{*}{$0.033 \pm 0.015$} & $\Xi^\prime_c(3/2^-)$ & $2.98 \pm 0.21$ & \multirow{2}{*}{$64 \pm 30$}
\\ \cline{9-10}
& & & & & & & & $\Xi^\prime_c(5/2^-)$ & $3.05 \pm 0.21$ &
\\ \cline{2-11}
& \multirow{2}{*}{$\Omega_c$} & \multirow{2}{*}{4.0} & \multirow{2}{*}{$0.52< T < 0.71$} & \multirow{2}{*}{$1.46 \pm 0.13$} & \multirow{2}{*}{$0.14 \pm 0.04$} & \multirow{2}{*}{$-2.89 \pm 0.47$} & \multirow{2}{*}{$0.025 \pm 0.014$} & $\Omega_c(3/2^-)$ & $3.27 \pm 0.17$ & \multirow{2}{*}{$50 \pm 27$}
\\ \cline{9-10}
& & & & & & & & $\Omega_c(5/2^-)$ & $3.32 \pm 0.17$ &
\\ \hline \hline
\end{tabular}
\label{tab:results1}
\end{center}
\end{table}

\begin{table}[hbt]
\begin{center}
\caption{We use the baryon multiplets $[\mathbf{\bar 3}_F, 1, 0, \lambda]$, $[\mathbf{\bar 3}_F, 0, 1, \rho]$, and $[\mathbf{\bar 3}_F, 2, 1, \rho]$ to fit the states $\Lambda_c(2595)$ ($J^P=1/2^-$), $\Lambda_c(2625)$ ($J^P=3/2^-$), $\Xi_c(2790)$ ($J^P=1/2^-$) and $\Xi_c(2815)$ ($J^P=3/2^-$), and use the multiplets $[\mathbf{6}_F, 1, 1, \lambda]$ and $[\mathbf{6}_F, 0, 1, \lambda]$ to fit the states $\Sigma_c(2800)$ ($J^P=?^?$), $\Xi_c(2980)$ ($J^P=?^?$) and $\Xi_c(3080)$ ($J^P=?^?$). The procedures are the same as before. Some fitting results are not so good and are summarized here. Sometimes the working region does not exist. For such cases, we choose the Borel Mass $T$ when the PC [defined in Eq.~(\ref{eq_pole})] is around 20\%, and show the CVG [defined in Eq.~(\ref{eq_convergence})] instead of working regions. We assume that free parameters $\omega_c$ in the same multiplet satisfy the relation $\omega_c(\Omega_c) - \omega_c(\Xi^\prime_c) = \omega_c(\Xi^\prime_c)- \omega_c(\Sigma_c)$, and use $\omega_c(\Omega_c)$ to evaluate the mass of $\Omega_c$.}
\begin{tabular}{c | c | c | c | c c | c c | c c | c}
\hline\hline
\multirow{2}{*}{Multiplets} & \multirow{2}{*}{B} & $\omega_c$ & Working region & $\overline{\Lambda}$ & $f$ & $K$ & $\Sigma$ & Baryons & Mass & Difference
\\ & & (GeV) & (GeV) & (GeV) & (GeV$^{4}$) & (GeV$^2$) & (GeV$^2$) & ($j^P$) & (GeV) & (MeV)
\\ \hline\hline
\multirow{4}{*}{$[\mathbf{\bar 3}_F, 1, 0, \lambda]$} & \multirow{2}{*}{$\Lambda_c$} & \multirow{2}{*}{2.9} &
\multirow{2}{*}{$-\left/\begin{array}{c}T=0.60\\{\rm CVG}=48\%\end{array}\right.$}
& \multirow{2}{*}{$0.96$} & \multirow{2}{*}{$0.03$} & \multirow{2}{*}{$-2.27$} & \multirow{2}{*}{$0.027$}
& $\Lambda_c(1/2^-)$ & $2.66$ & \multirow{2}{*}{$32$}
\\ \cline{9-10}
& & & & & & & & $\Lambda_c(3/2^-)$ & $2.69$ &
\\ \cline{2-11}
& \multirow{2}{*}{$\Xi_c$} & \multirow{2}{*}{3.1} &
\multirow{2}{*}{$-\left/\begin{array}{c}T=0.63\\{\rm CVG}=47\%\end{array}\right.$}
& \multirow{2}{*}{$1.06$} & \multirow{2}{*}{$0.04$} & \multirow{2}{*}{$-2.46$} & \multirow{2}{*}{$0.025$} & $\Xi_c(1/2^-)$ & $2.79$ & \multirow{2}{*}{$29$}
\\ \cline{9-10}
& & & & & & & & $\Xi_c(3/2^-)$ & $2.82$ &
\\ \hline
\multirow{3}{*}{$[\mathbf{\bar 3}_F, 0, 1, \rho]$} & $\Lambda_c$ & 3.5 &
{$-\left/\begin{array}{c}T=0.73\\{\rm CVG}=50\%\end{array}\right.$}
& $0.99$ & $0.03$ & $-1.77$ & $0$ & $\Lambda_c(1/2^-)$ & $2.61$ & $-$
\\ \cline{2-11}
& $\Xi_c$ & 3.1 &
{$-\left/\begin{array}{c}T=0.76\\{\rm CVG}=77\%\end{array}\right.$}
& $1.18$ & $0.04$ & $-2.09$ & $0$ & $\Xi_c(1/2^-)$ & $2.87$ & $-$
\\ \hline
\multirow{4}{*}{$[\mathbf{\bar 3}_F, 2, 1, \rho]$} & \multirow{2}{*}{$\Lambda_c$} & \multirow{2}{*}{3.6} &
\multirow{2}{*}{$-\left/\begin{array}{c}T=0.59\\{\rm CVG}=43\%\end{array}\right.$}
& \multirow{2}{*}{$1.34$} & \multirow{2}{*}{$0.08$} & \multirow{2}{*}{$-0.23$} & \multirow{2}{*}{$0.050$} & $\Lambda_c(3/2^-)$ & $2.60$ & \multirow{2}{*}{$98$}
\\ \cline{9-10}
& & & & & & & & $\Lambda_c(5/2^-)$ & $2.70$ &
\\ \cline{2-11}
& \multirow{2}{*}{$\Xi_c$} & \multirow{2}{*}{4.0} &
\multirow{2}{*}{$-\left/\begin{array}{c}T=0.66\\{\rm CVG}=34\%\end{array}\right.$}
& \multirow{2}{*}{$1.51$} & \multirow{2}{*}{$0.13$} & \multirow{2}{*}{$-0.51$} & \multirow{2}{*}{$0.040$} & $\Xi_c(3/2^-)$ & $2.84$ & \multirow{2}{*}{$79$}
\\ \cline{9-10}
& & & & & & & & $\Xi_c(5/2^-)$ & $2.92$ &
\\ \hline
\multirow{3}{*}{$[\mathbf{6}_F, 0, 1, \lambda]$} & $\Sigma_c$ & 2.9 &
{$-\left/\begin{array}{c}T=0.57\\{\rm CVG}=36\%\end{array}\right.$}
& $1.10$ & $0.05$ & $-2.28$ & $0$ & $\Sigma_c(1/2^-)$ & $2.82$ & $-$
\\ \cline{2-11}
& $\Xi^\prime_c$ & 3.4 & {$0.54< T < 0.61$} & $1.30 \pm 0.11$ & $0.07 \pm 0.02$ & $-2.23 \pm 0.42$ & $0$ & $\Xi^\prime_c(1/2^-)$ & $3.01 \pm 0.14$ & $-$
\\ \cline{2-11}
& $\Omega_c$ & 3.9 & {$0.54< T < 0.66$} & $1.50 \pm 0.09$ & $0.11 \pm 0.03$ & $-2.39 \pm 0.65$ & $0$ & $\Omega_c(1/2^-)$ & $3.25 \pm 0.16$ & $-$
\\ \hline
\multirow{6}{*}{$[\mathbf{6}_F, 1, 1, \lambda]$} & \multirow{2}{*}{$\Sigma_c$} & \multirow{2}{*}{3.5} & \multirow{2}{*}{$0.64< T < 0.67$} & \multirow{2}{*}{$1.07 \pm 0.11$} & \multirow{2}{*}{$0.05 \pm 0.01$} & \multirow{2}{*}{$-2.40 \pm 0.28$} & \multirow{2}{*}{$\begin{array}{c}0.032 \\ \pm0.008\end{array}$}
& $\Sigma_c(1/2^-)$ & $2.79 \pm 0.13$ & \multirow{2}{*}{$38 \pm 9$}
\\ \cline{9-10}
& & & & & & & & $\Sigma_c(3/2^-)$ & $2.82 \pm 0.13$ &
\\ \cline{2-11}
& \multirow{2}{*}{$\Xi^\prime_c$} & \multirow{2}{*}{3.5} &
\multirow{2}{*}{$-\left/\begin{array}{c}T=0.70\\{\rm CVG}=40\%\end{array}\right.$}
& \multirow{2}{*}{$1.18$} & \multirow{2}{*}{$0.06$} & \multirow{2}{*}{$-2.81$} & \multirow{2}{*}{$0.032$} & $\Xi^\prime_c(1/2^-)$ & $2.98$ & \multirow{2}{*}{$37$}
\\ \cline{9-10}
& & & & & & & & $\Xi^\prime_c(3/2^-)$ & $3.02$ &
\\ \cline{2-11}
& \multirow{2}{*}{$\Omega_c$} & \multirow{2}{*}{3.5} &
\multirow{2}{*}{$-\left/\begin{array}{c}T=0.71\\{\rm CVG}=43\%\end{array}\right.$}
& \multirow{2}{*}{$1.27$} & \multirow{2}{*}{$0.08$} & \multirow{2}{*}{$-3.01$} & \multirow{2}{*}{$0.029$} & $\Omega_c(1/2^-)$ & $3.11$ & \multirow{2}{*}{$34$}
\\ \cline{9-10}
& & & & & & & & $\Omega_c(3/2^-)$ & $3.15$ &
\\ \hline \hline
\end{tabular}
\label{tab:results2}
\end{center}
\end{table}

During the numerical analyses, we find that all the curves of CVF, PC, $\overline{\Lambda}_{F,j_l,s_l,\rho/\lambda}$, $K_{F,j_l,s_l,\rho/\lambda}$, and $\Sigma_{F,j_l,s_l,\rho/\lambda}$ behave similarly to those shown in Figs.~\ref{fig:pole}-\ref{fig:next}, although sometimes the working region does not exist. For such cases, we choose the Borel Mass $T$ when the PC [defined in Eq.~(\ref{eq_pole})] is around 20\%, and show the CVG [defined in Eq.~(\ref{eq_convergence})] in Table~\ref{tab:results2} instead of working regions. The sum rules in such cases are not good, suggesting the relevant states are not dominated by the components related to the currents used. Moreover, when CVG is larger than 50\%, the high-order power corrections are already larger than the perturbation term. The sum rules in such cases are bad, suggesting the relevant states should not significantly contain the components related to the currents used. In these cases we do not evaluate the error bars for simplicity.

Our conclusions are the following:
\begin{enumerate}

\item The baryon doublet $[\mathbf{\bar 3}_F, 1, 1, \rho]$ contains $\Lambda_c(1/2^-,3/2^-)$ and $\Xi_c(1/2^-,3/2^-)$; see Table~\ref{tab:results1}. We use them to perform QCD sum rule analyses, and the obtained masses as well as their splittings are well consistent with the observed states $\Lambda_c(2595)$ ($J^P=1/2^-$), $\Lambda_c(2625)$ ($J^P=3/2^-$), $\Xi_c(2790)$ ($J^P=1/2^-$), and $\Xi_c(2815)$ ($J^P=3/2^-$)~\cite{pdg}. Our results suggest that these states contain $[\mathbf{\bar 3}_F, 1, 1, \rho]$ components. Indeed, these states can be well described by the heavy doublet $[\mathbf{\bar 3}_F, 1, 1, \rho]$, and they complete two $SU(3)$ $\mathbf{\bar 3}_F$ multiplets of $J^P=1/2^-$ and $3/2^-$.

\item The baryon doublet $[\mathbf{6}_F, 1, 0, \rho]$ contains $\Sigma_c(1/2^-,3/2^-)$, $\Xi^\prime_c(1/2^-,3/2^-)$, and $\Omega_c(1/2^-,3/2^-)$; see Table~\ref{tab:results1}. We use them to perform QCD sum rule analyses, and the obtained results are consistent with the observed states $\Sigma_c(2800)$ ($J^P=?^?$) and $\Xi_c(2980)$ ($J^P=?^?$)~\cite{pdg}. As an example, we show the variations of $\Lambda_{\Xi^\prime_c,1,0,\rho}$, $f_{\Xi^\prime_c,1,0,\rho}$, $K_{\Xi^\prime_c,1,0,\rho}$ and $\Sigma_{\Xi^\prime_c,1,0,\rho}$ with respect to the Borel mass $T$ in Fig.~\ref{fig:Xi10}, when $J_{1/2,-,\Xi^\prime_c,1,0,\rho}$ is used. Our results suggest that these states contain $[\mathbf{6}_F, 1, 0, \rho]$ components. Our results also suggest that there are two $\Sigma_c(2800)$ states of $J^P = 1/2^-$ and $3/2^-$, whose mass splitting is $14 \pm 7$ MeV; there are two $\Xi_c(2980)$ states, whose mass splitting is $12 \pm 7$ MeV; there are also two $\Omega_c$ states of $J^P = 1/2^-$ and $3/2^-$, whose masses are around $3.25\pm0.20$ GeV with mass splitting $10 \pm 6$ MeV.

\begin{figure}[hbt]
\begin{center}
\scalebox{0.598}{\includegraphics{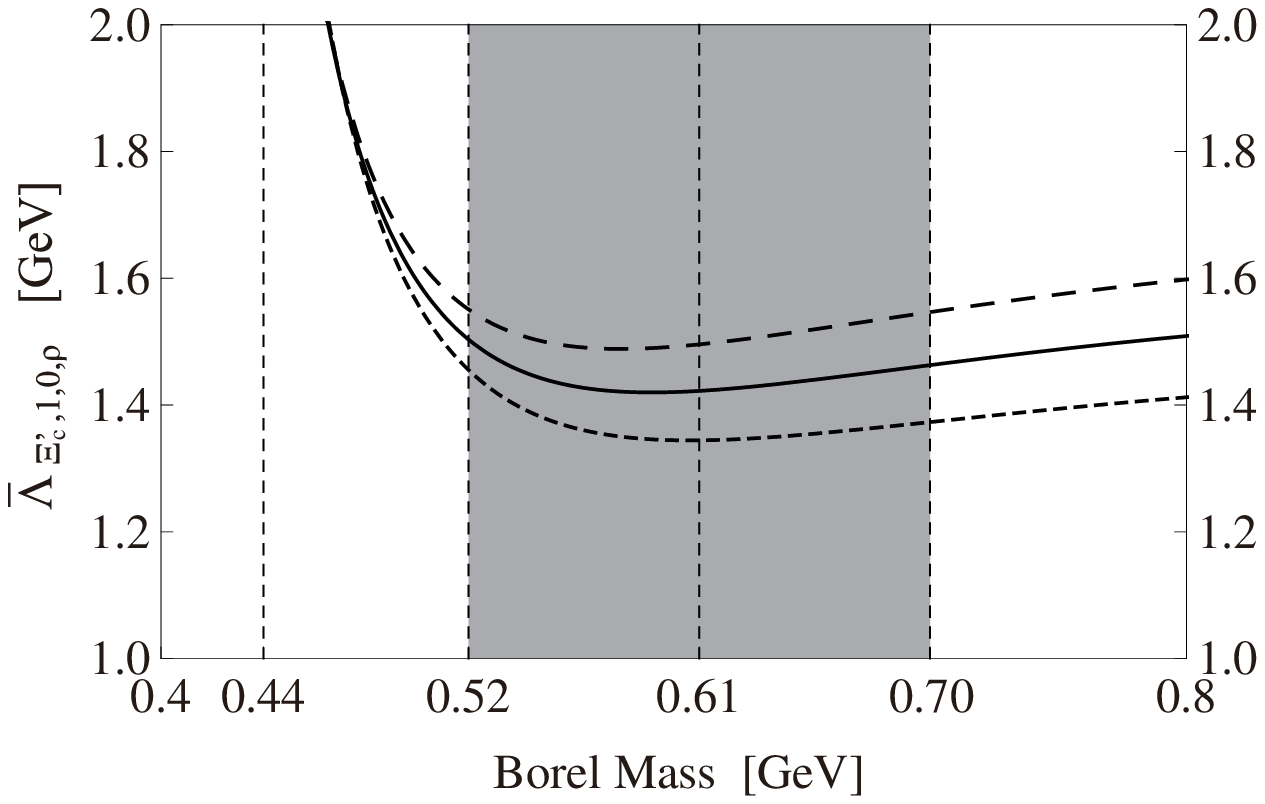}}
\scalebox{0.6}{\includegraphics{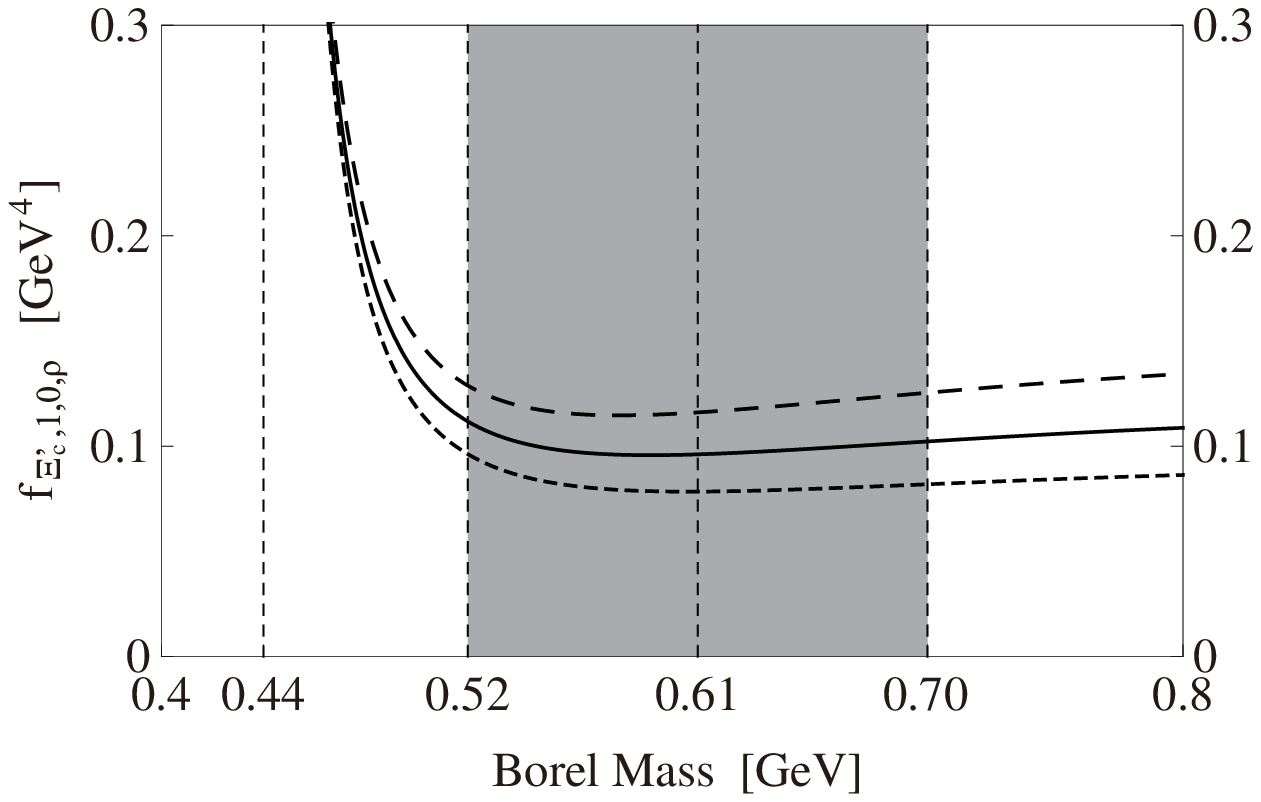}}
\\ \scalebox{0.612}{\includegraphics{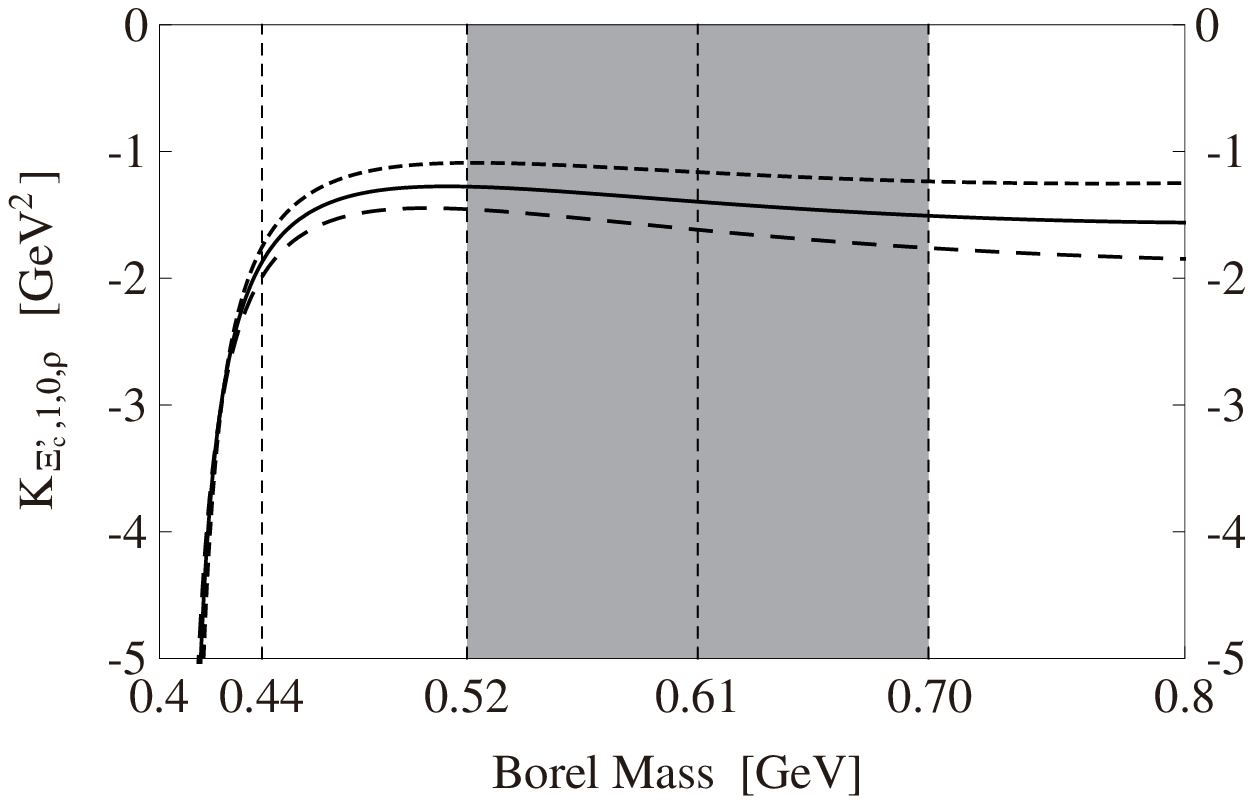}}
\scalebox{0.58}{\includegraphics{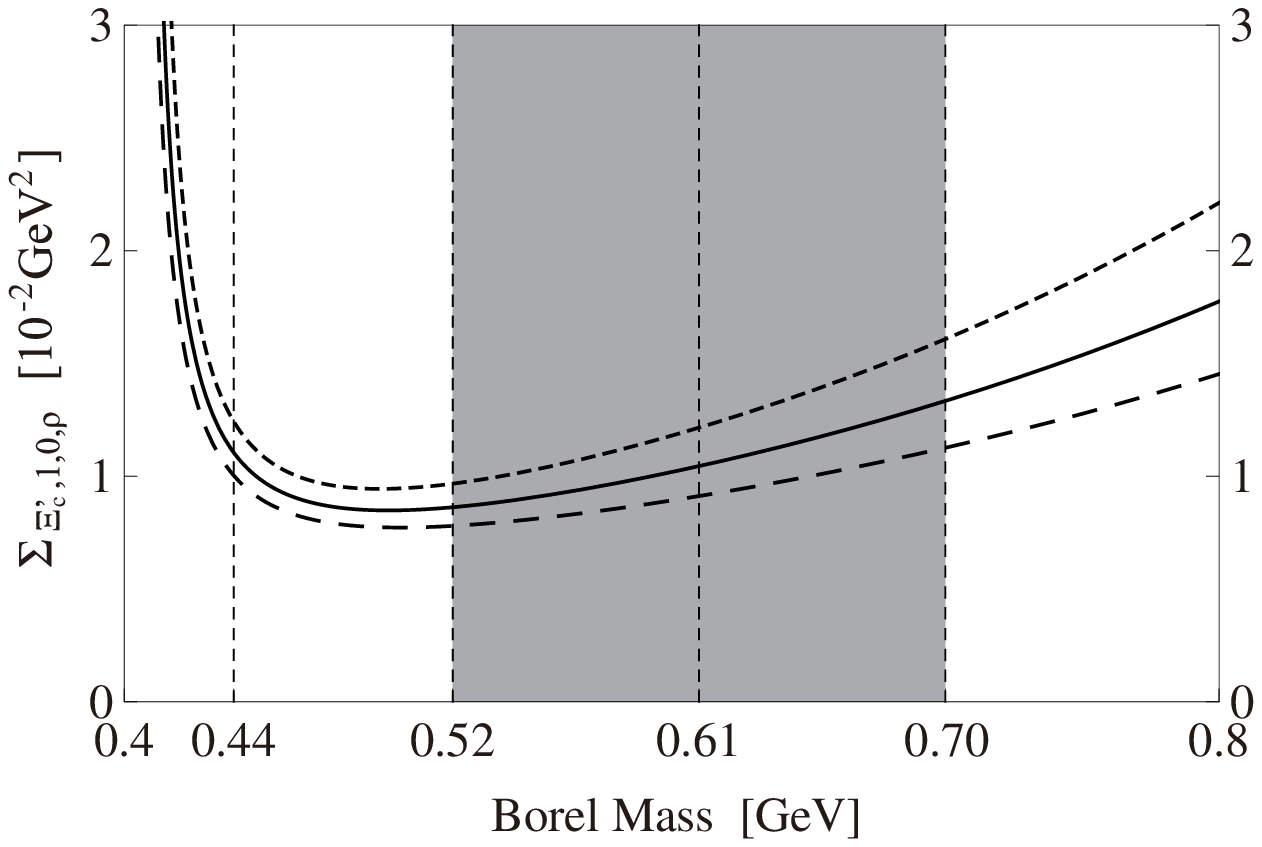}} \caption{The variations
of $\overline{\Lambda}_{\Xi^\prime_c,1,0,\rho}$, $f_{\Xi^\prime_c,1,0,\rho}$, $K_{\Xi^\prime_c,1,0,\rho}$, and $\Sigma_{\Xi^\prime_c,1,0,\rho}$ with respect to the Borel mass
$T$, when $J_{1/2,-,\Xi^\prime_c,1,0,\rho}$ is used. We use the working region $0.52$ GeV $< T < 0.70$
GeV, and obtain the short-dashed, solid, and
long-dashed curves by fixing $\omega_c = 3.7$, 3.9, and
4.1 GeV, respectively.}
\label{fig:Xi10}
\end{center}
\end{figure}

\item The baryon doublet $[\mathbf{6}_F, 2, 1, \lambda]$ contains $\Sigma_c(3/2^-,5/2^-)$, $\Xi^\prime_c(3/2^-,5/2^-)$, and $\Omega_c(3/2^-,5/2^-)$; see Table~\ref{tab:results1}. We use them to perform QCD sum rule analyses, and the obtained results are consistent with the observed states $\Sigma_c(2800)$ ($J^P=?^?$), $\Xi_c(2980)$ ($J^P=?^?$), and $\Xi_c(3080)$ ($J^P=?^?$)~\cite{pdg}. As an example, we show the variations of $\overline{\Lambda}_{\Xi^\prime_c,2,1,\lambda}$, $f_{\Xi^\prime_c,2,1,\lambda}$, $K_{\Xi^\prime_c,2,1,\lambda}$, and $\Sigma_{\Xi^\prime_c,2,1,\lambda}$ with respect to the Borel mass $T$ in Fig.~\ref{fig:Xi21}, when $J_{3/2,-,\Xi^\prime_c,2,1,\lambda}$ is used. Particularly, we obtain a mass splitting $64 \pm 30$ MeV between two $\Xi^\prime_c$ states, which is not far from the mass difference 109 MeV between $\Xi_c(2980)$ and $\Xi_c(3080)$, suggesting that $\Xi_c(3080)$ may be the $5/2^-$ partner of $\Xi_c(2980)$. If this is the case, $\Sigma_c(2800)$ and $\Omega_c(3/2^-)$ may also have $5/2^-$ partners, whose masses are $85 \pm 23$ MeV and $50 \pm 27$ MeV larger. However, we do not draw firm conclusions here because there are many excited $\Xi_c$ states theoretically, and $\Xi_c(3080)$ may belong to other baryon multiplets and have different quantum numbers other than $5/2^-$.

\begin{figure}[hbt]
\begin{center}
\scalebox{0.597}{\includegraphics{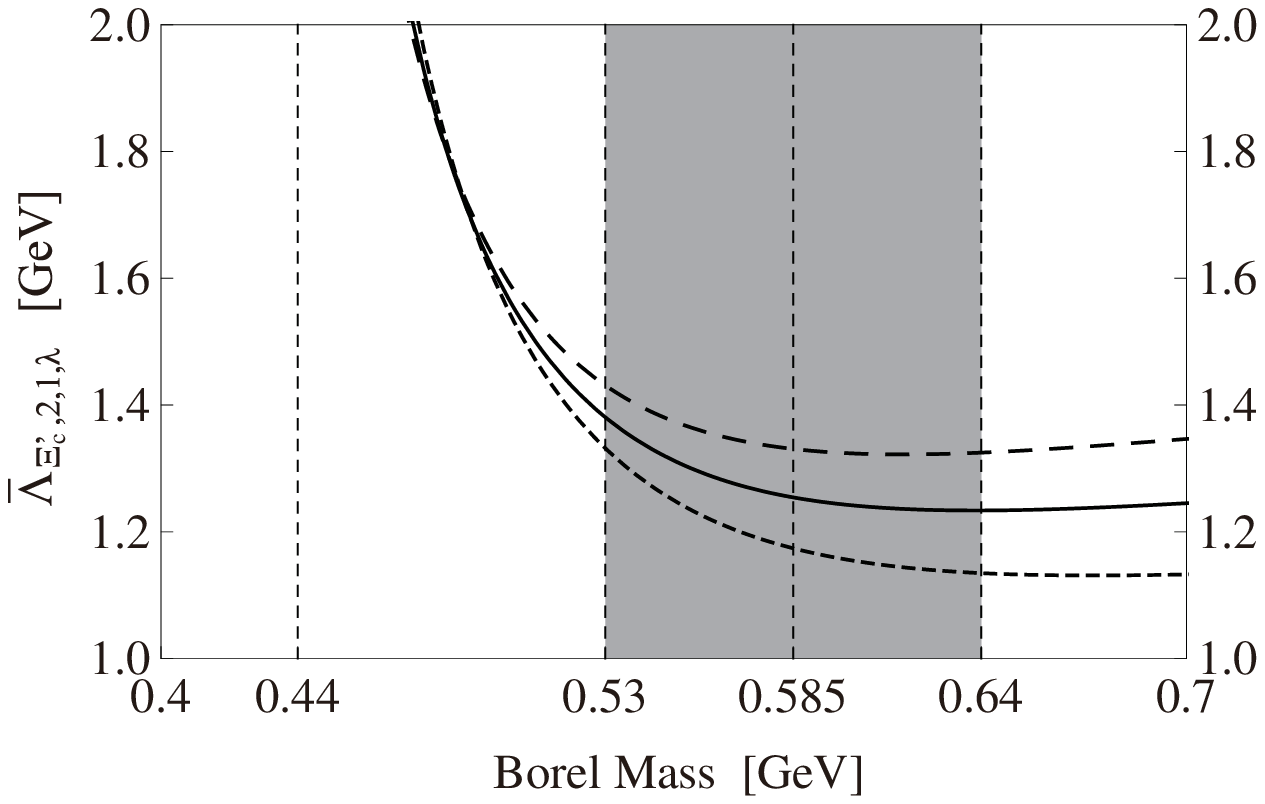}}
\scalebox{0.6}{\includegraphics{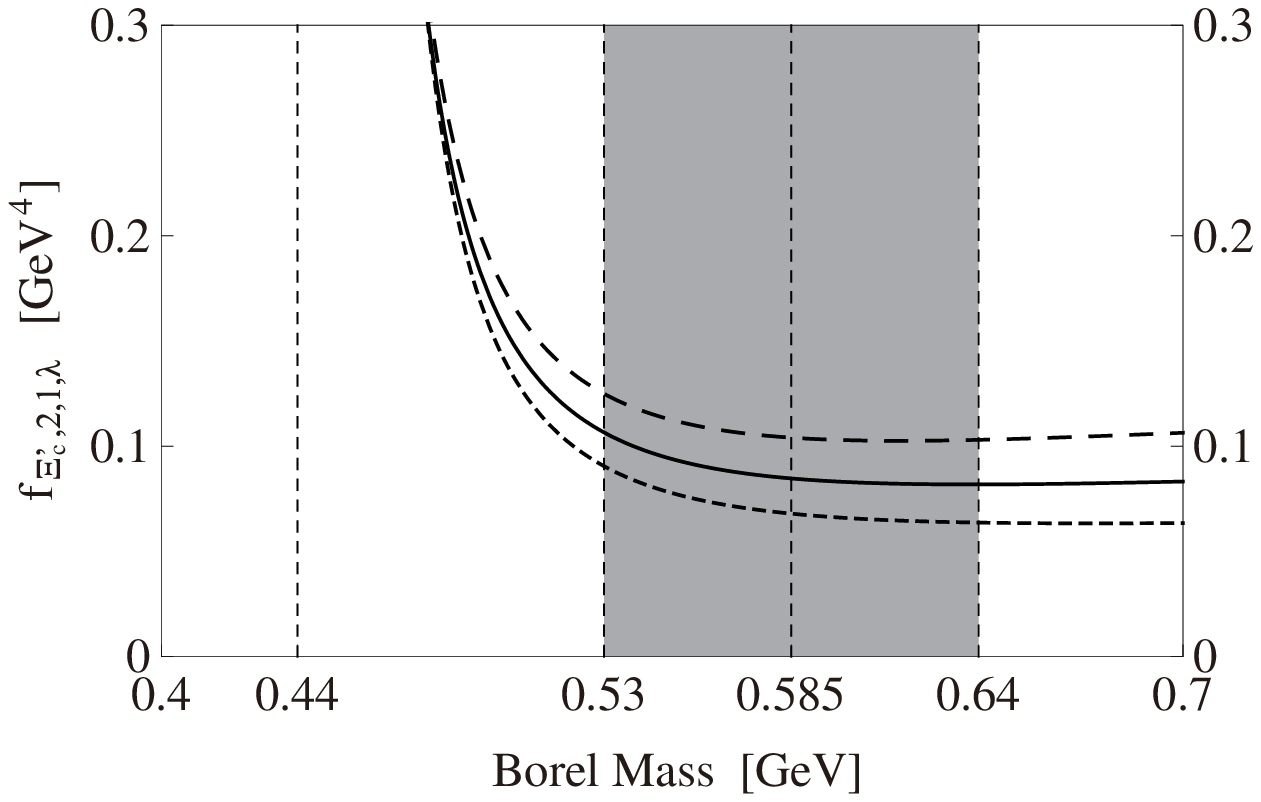}}
\\ \scalebox{0.6}{\includegraphics{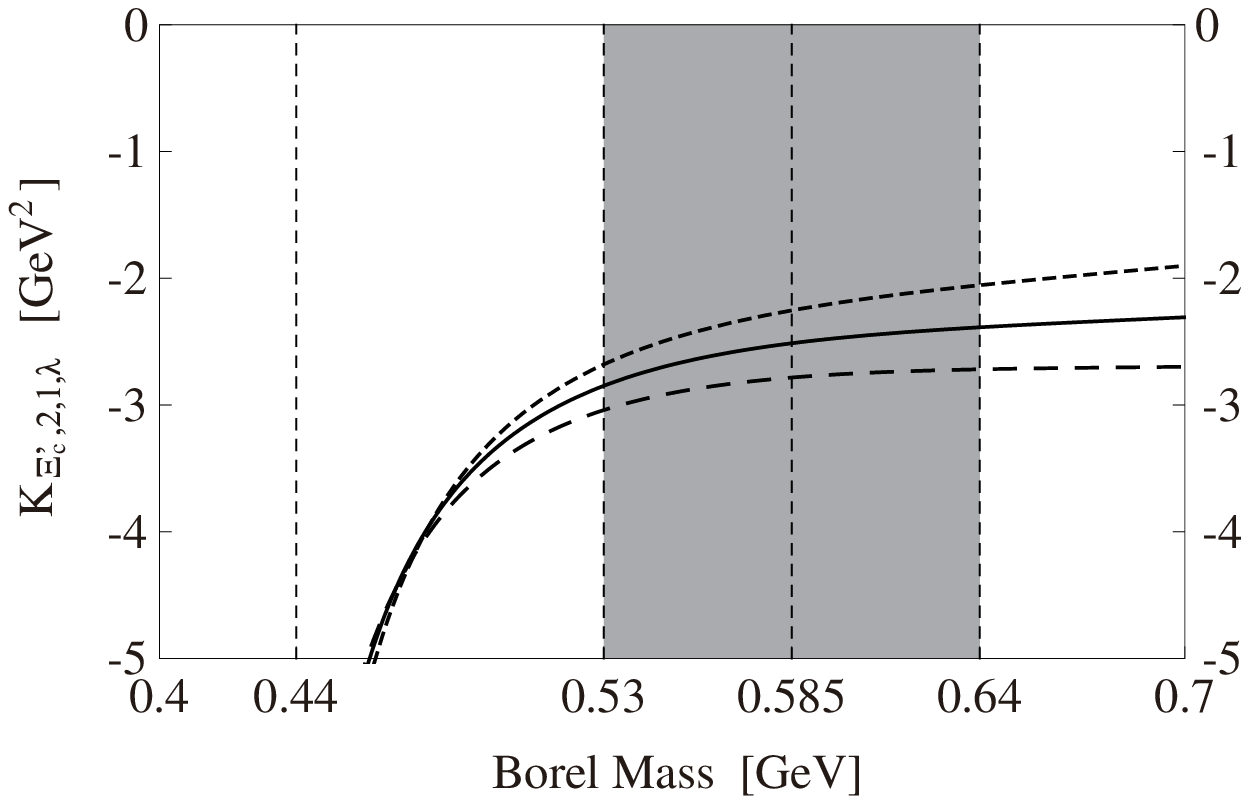}}
\scalebox{0.594}{\includegraphics{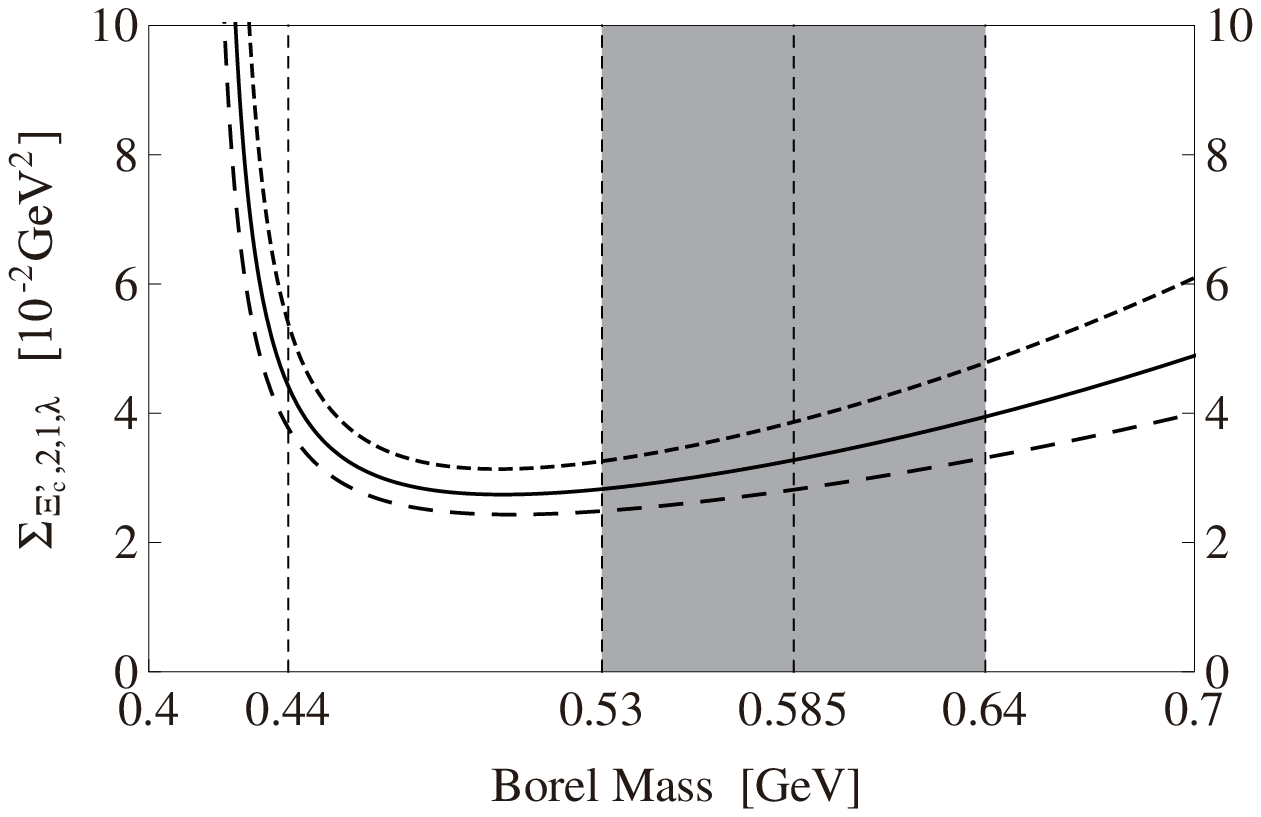}} \caption{The variations
of $\overline{\Lambda}_{\Xi^\prime_c,2,1,\lambda}$, $f_{\Xi^\prime_c,2,1,\lambda}$, $K_{\Xi^\prime_c,2,1,\lambda}$, and $\Sigma_{\Xi^\prime_c,2,1,\lambda}$ with respect to the Borel mass
$T$, when $J_{3/2,-,\Xi^\prime_c,2,1,\lambda}$ is used. We work in the region $0.53$ GeV $< T < 0.64$
GeV, and obtain the short-dashed, solid, and
long-dashed curves by fixing $\omega_c = 3.3$, 3.5, and
3.7 GeV, respectively.}
\label{fig:Xi21}
\end{center}
\end{figure}

\item Other not so good results are listed in Table~\ref{tab:results2} for completeness, where not all the working regions exist. We do not use them to draw any conclusion, but just note that the present sum rule analysis finds a better working window by the $\rho$ mode, while the $\lambda$ mode also provides reasonable results consistent with the experiments: a) the mass splittings obtained by using the baryon doublet $[\mathbf{\bar 3}_F, 1, 0, \lambda]$ are very well consistent with the observed states $\Lambda_c(2595)$, $\Lambda_c(2625)$, $\Xi_c(2790)$, and $\Xi_c(2815)$~\cite{pdg}, suggesting that these states may contain $[\mathbf{\bar 3}_F, 1, 0, \lambda]$ components; b) the sum rule to calculate the $\Omega_c(1/2^-)$ mass using the baryon doublet $[\mathbf{6}_F, 0, 1, \lambda]$ does have a working region, and the obtained result is around $3.25\pm0.16$ GeV, supporting the above analyses.

\end{enumerate}

Summarizing all these results, we have studied the $P$-wave charmed baryons using the method of QCD sum rule in the framework of HQET.
We have calculated their masses up to the $\mathcal{O}(1/m_Q)$ order, the results of which have large theoretical uncertainties.
We have also calculated their mass splittings within the same doublet, the results of which are reproduced quite well with much less theoretical uncertainty.
Our results suggest that the four observed states $\Lambda_c(2595)$ ($J^P=1/2^-$), $\Lambda_c(2625)$ ($J^P=3/2^-$), $\Xi_c(2790)$ ($J^P=1/2^-$) and $\Xi_c(2815)$ ($J^P=3/2^-$)
can be well described by the heavy doublet $[\mathbf{\bar 3}_F, 1, 1, \rho]$ and they complete two $SU(3)$ $\mathbf{\bar 3}_F$ multiplets of $J^P=1/2^-$ and $3/2^-$.
The $SU(3)$ $\mathbf{6}_F$ multiplets are more complicated. Our results suggest that $\Sigma_c(2800)$ ($J^P=?^?$) and $\Xi_c(2980)$ ($J^P=?^?$) belong to these multiplets, but there are two $\Sigma_c(2800)$ states of $J^P = 1/2^-$ and $3/2^-$ whose mass splitting is $14 \pm 7$ MeV, and two $\Xi_c(2980)$ states whose mass splitting is $12 \pm 7$ MeV. They have two $\Omega_c$ partners of $J^P = 1/2^-$ and $3/2^-$, whose masses are around $3.25\pm0.20$ GeV with mass splitting $10 \pm 6$ MeV. All of them together complete two  $SU(3)$ $\mathbf{6}_F$ multiplets of $J^P=1/2^-$ and $3/2^-$. They may have three $J^P=5/2^-$ partners. $\Xi_c(3080)$ ($J^P=?^?$) may be one of them, and the other two are $\Sigma_c(5/2^-)$ and $\Omega_c(5/2^-)$, whose masses are $85 \pm 23$ and $50 \pm 27$ MeV larger.

To end our paper, we note that in a nonrelativistic model of attractive potential of the form (distance)$^n$,
the excitation energies of the $\lambda$ mode should appear lower than the corresponding one
of the $\rho$ mode for a positive power $n$,
while this order interchanges for a negative $n$  ($n$ must satisfy $n \geq -1$ for stable solutions to exist).
Thus, our present analysis implies that further investigations would be needed to clarify the nature of heavy baryon excitations, while the conventional quark model results seem reasonable with lower $\lambda$ modes~\cite{Copley:1979wj}.
Another related
subject is the $P$-wave bottom baryons, which can be similarly studied, and we are now doing these analyses.
Particularly, the mass difference between $\Sigma_b({1/2}^-)$ and $\Sigma_b({3/2}^-)$ for the baryon
doublet $[\mathbf{\bar 3}_F, 1, 1, \rho]$ can be roughly estimated, and the result is around (see Table~\ref{tab:results1})
\begin{eqnarray}
m_{\Sigma_b({3/2}^-)} - m_{\Sigma_b({1/2}^-)} \approx C_{mag} \times {1 \over 4m_b} \times {6} \times [ 0.042 \mbox{ GeV}^2 ] \approx 12 ~{\rm MeV} \, ,
\end{eqnarray}
where we have used $C_{mag} \approx 0.8$~\cite{Dai:1996qx,Dai:2003yg} and $m_b = 4.18$ GeV~\cite{pdg}. It is consistent with the mass difference of $\Sigma_b(5912)$ ($J^P=1/2^-$) and $\Sigma_b(5920)$ ($J^P=3/2^-$)~\cite{pdg}.

\section*{ACKNOWLEDGMENTS}

We thank Cheng-Ping Shen for useful discussions.
This project is supported by the National Natural Science Foundation
of China under Grants No. 11205011, No. 11475015, No. 11375024, No. 11222547, No.
11175073, and No. 11261130311, the Ministry of
Education of China (SRFDP under Grant No. 20120211110002 and the
Fundamental Research Funds for the Central Universities), and the
Fok Ying-Tong Education Foundation (Grant No. 131006).

\appendix

\section{Other Sum Rules}
\label{sec:others}

In this appendix we show the sum rules for other currents with different quark contents:
\begin{eqnarray}
&& \Pi_{1/2,-,\Sigma_c,1,0,\rho} = f_{\Sigma_c,1,0,\rho}^2 e^{-2 \overline{\Lambda}_{\Sigma_c,1,0,\rho} / T}
\\ \quad && =
\int_{0}^{\omega_c} [ \frac{1}{3584\pi^4}\omega^7 ]e^{-\omega/T}d\omega - \frac{3\langle g_s^2 GG \rangle}{512 \pi^4} T^4 + \frac{\langle g_s \bar q \sigma Gq \rangle \langle \bar q q \rangle}{4} - \frac{\langle g_s \bar q \sigma G q\rangle^2}{16} {1 \over T^2} \, ,
\nonumber \\
&& f_{\Sigma_c,1,0,\rho}^2 K_{\Sigma_c,1,0,\rho} e^{-2 \overline{\Lambda}_{\Sigma_c,1,0,\rho} / T}
\nonumber\\ \quad && =\int_{0}^{\omega_c}[- \frac{1}{17920\pi^4}\omega^9] e^{-\omega/T} d\omega + \frac{5\langle g_s^2 GG \rangle}{256 \pi^4} T^6- \frac{\langle g_s \bar q \sigma G q\rangle^2}{16}+\frac{\langle g_s \bar q \sigma Gq \rangle \langle \bar q q \rangle\langle g_s^2 GG \rangle}{64}{1 \over T^2}
\nonumber\\ \quad &&\quad
- \frac{\langle g_s \bar q \sigma Gq \rangle ^2\langle g_s^2 GG \rangle}{256}{1 \over T^4} \, ,
\nonumber \\
\nonumber && f_{\Sigma_c,1,0,\rho}^2 \Sigma_{\Sigma_c,1,0,\rho} e^{-2 \overline{\Lambda}_{\Sigma_c,1,0,\rho} / T}
={{\langle g_s^2 GG \rangle}\over 64 \pi^4} T^6 \, .
\end{eqnarray}

\begin{eqnarray}
&& \Pi_{1/2,-,\Xi^\prime_c,1,0,\rho}(\omega_c, T) = f_{\Xi^\prime_c,1,0,\rho}^2 e^{-2 \overline{\Lambda}_{\Xi^\prime_c,1,0,\rho} / T}
\\ \quad && =
\int_{2m_s}^{\omega_c} [ \frac{1}{3584\pi^4}\omega^7-\frac{3m_s^2}{320\pi^4}\omega^5 ]e^{-\omega/T}d\omega - \frac{3m_s\langle \bar q q \rangle}{8 \pi^2} T^4
+ \frac{9m_s\langle \bar s s \rangle}{16 \pi^2} T^4 -\frac{3\langle g_s^2 GG \rangle}{512 \pi^4} T^4
\nonumber\\ \quad &&\quad
+\frac{3m_s^2\langle g_s^2 GG \rangle}{256 \pi^4} T^2+\frac{\langle g_s\bar q \sigma Gq \rangle\langle \bar s s \rangle}{8}
+\frac{\langle g_s\bar s \sigma Gs \rangle\langle \bar q q \rangle}{8}-\frac{m_s\langle \bar s s \rangle\langle g_s^2 GG \rangle}{128 \pi^2}- \frac{\langle g_s \bar q \sigma Gq\rangle\langle g_s\bar s \sigma Gs \rangle}{16} {1 \over T^2} \, ,
\nonumber \\
\nonumber && f_{\Xi^\prime_c,1,0,\rho}^2 K_{\Xi^\prime_c,1,0,\rho} e^{-2 \overline{\Lambda}_{\Xi^\prime_c,1,0,\rho} / T}
\\ \quad && =\int_{2m_s}^{\omega_c} [ -\frac{1}{17920\pi^4}\omega^9+\frac{47m_s^2}{17920\pi^4}\omega^7 ]e^{-\omega/T}d\omega +\frac{15m_s\langle \bar q q \rangle}{4 \pi^2} T^6-\frac{6m_s\langle \bar s s \rangle}{\pi^2} T^6+\frac{5\langle g_s^2 GG \rangle}{256\pi^4} T^6
\nonumber\\ \quad &&\quad
-\frac{3m_s\langle g_s \bar q \sigma Gq \rangle}{4 \pi^2} T^4-\frac{3m_s^2\langle g_s^2 GG \rangle}{64 \pi^4} T^4-\frac{\langle g_s \bar q \sigma Gq \rangle\langle g_s \bar s \sigma Gs \rangle}{16}-\frac{3m_s\langle \bar q q \rangle\langle g_s^2 GG \rangle}{128 \pi^2} T^2+\frac{m_s\langle \bar s s \rangle\langle g_s^2 GG \rangle}{128 \pi^2} T^2
\nonumber\\ \quad &&\quad
+\frac{\langle \bar q q \rangle \langle g_s \bar s \sigma Gs \rangle\langle g_s^2 GG \rangle}{128}{1 \over T^2}+\frac{\langle g_s \bar q \sigma Gq \rangle\langle \bar s s \rangle\langle g_s^2 GG \rangle}{128}{1 \over T^2}-\frac{\langle g_s \bar q \sigma Gq \rangle\langle g_s \bar s \sigma Gs \rangle\langle g_s^2 GG \rangle}{256}{1 \over T^4} \, ,
\nonumber \\
\nonumber && f_{\Xi^\prime_c,1,0,\rho}^2 \Sigma_{\Xi^\prime_c,1,0,\rho} e^{-2 \overline{\Lambda}_{\Xi^\prime_c,1,0,\rho} / T}
=\frac{\langle g_s^2 GG \rangle}{64 \pi^4} T^6-\frac{m_s^2\langle g_s^2 GG \rangle}{128 \pi^4} T^4+\frac{m_s\langle \bar s s \rangle\langle g_s^2 GG \rangle}{192 \pi^2} T^2 \, .
\end{eqnarray}

\begin{eqnarray}
&& \Pi_{1/2,-,\Omega_c,1,0,\rho}= f_{\Omega_c,1,0,\rho}^2 e^{-2 \overline{\Lambda}_{\Omega_c,1,0,\rho} / T}
\\ \quad && =
\int_{4m_s}^{\omega_c} [ \frac{1}{3584\pi^4}\omega^7-  \frac{9m_s^2}{640\pi^4}\omega^5+\frac{9m_s^4}{64\pi^4}\omega^3]e^{-\omega/T}d\omega +\frac{3m_s\langle \bar s s \rangle}{8 \pi^2} T^4- \frac{3\langle g_s^2 GG \rangle}{512 \pi^4} T^4 -\frac{3m_s^3\langle \bar s s \rangle}{4\pi^2} T^2 +\frac{3m_s^2\langle g_s^2 GG \rangle}{128\pi^2} T^2
\nonumber\\ \quad &&\quad
+\frac{m_s^2\langle \bar s s \rangle^2}{8}+\frac{\langle g_s \bar s \sigma Gs \rangle \langle \bar s s \rangle}{4}-\frac{m_s\langle \bar s s \rangle\langle g_s^2 GG \rangle}{64 \pi^2}-\frac{\langle g_s \bar s \sigma Gs \rangle^2}{16} {1 \over T^2} \, ,
\nonumber \\
&& f_{\Omega_c,1,0,\rho}^2 K_{\Omega_c,1,0,\rho} e^{-2 \overline{\Lambda}_{\Omega_c,1,0,\rho} / T}
\nonumber\\ \quad && =\int_{4m_s}^{\omega_c} [ -\frac{1}{17920\pi^4}\omega^9+  \frac{9m_s^2}{2240\pi^4}\omega^7]e^{-\omega/T}d\omega -\frac{9m_s\langle \bar s s \rangle}{2 \pi^2} T^6+\frac{5\langle g_s^2 GG \rangle}{256\pi^4} T^6-\frac{3m_s\langle g_s \bar s \sigma Gs \rangle}{ 2\pi^2}T^4-\frac{15m_s^2\langle g_s^2 GG \rangle}{ 256\pi^4}T^4
\nonumber\\ \quad &&\quad
-\frac{\langle g_s \bar s \sigma Gs \rangle^2}{16}
-\frac{m_s\langle \bar s s \rangle\langle g_s^2 GG \rangle}{ 32\pi^2}T^2+\frac{\langle \bar s s \rangle\langle g_s \bar s \sigma Gs \rangle\langle g_s^2 GG \rangle}{64}{1 \over T^2}-\frac{m_s^2\langle \bar s s \rangle^2\langle g_s^2 GG \rangle}{128}{1 \over T^2}-\frac{\langle g_s \bar s \sigma Gs \rangle^2\langle g_s^2 GG \rangle}{256}{1 \over T^4}\, ,
\nonumber \\
\nonumber && f_{\Omega_c,1,0,\rho}^2 \Sigma_{\Omega_c,1,0,\rho} e^{-2 \overline{\Lambda}_{\Omega_c,1,0,\rho} / T}
=\frac{\langle g_s^2 GG \rangle}{64\pi^4} T^6-\frac{m_s^2\langle g_s^2 GG \rangle}{64 \pi^4} T^4+\frac{m_s\langle \bar s s \rangle\langle g_s^2 GG \rangle}{96\pi^2} T^2 \, .
\end{eqnarray}

\begin{eqnarray}
&& \Pi_{1/2,-,\Lambda_c,0,1,\rho}= f_{\Lambda_c,0,1,\rho}^2 e^{-2 \overline{\Lambda}_{\Lambda_c,0,1,\rho} / T}
\\ \quad && =
\int_{0}^{\omega_c} [ \frac{1}{17920\pi^4}\omega^7 ]e^{-\omega/T}d\omega +\frac{\langle g_s^2 GG \rangle}{512 \pi^4} T^4+\frac{\langle g_s \bar q \sigma Gq \rangle \langle \bar q q \rangle}{4}-\frac{\langle g_s \bar q \sigma Gq \rangle^2}{16} {1 \over T^2} \, ,
\nonumber \\
&& f_{\Lambda_c,0,1,\rho}^2 K_{\Lambda_c,0,1,\rho} e^{-2 \overline{\Lambda}_{\Lambda_c,0,1,\rho} / T}
\nonumber\\ \quad && =\int_{0}^{\omega_c} [ -\frac{1}{80640\pi^4}\omega^9 ]e^{-\omega/T}d\omega - \frac{13\langle g_s^2 GG \rangle}{256 \pi^4} T^6-\frac{\langle g_s \bar q \sigma Gq \rangle^2}{16}
+\frac{\langle \bar q q \rangle \langle g_s \bar q \sigma Gq \rangle\langle g_s^2 GG \rangle}{64}{1 \over T^2}
\nonumber\\ \quad &&\quad
- \frac{\langle g_s \bar q \sigma Gq \rangle^2\langle g_s^2 GG \rangle}{256}{1 \over T^4} \, ,
\nonumber \\
\nonumber && f_{\Lambda_c,0,1,\rho}^2 \Sigma_{\Lambda_c,0,1,\rho} e^{-2 \overline{\Lambda}_{\Lambda_c,0,1,\rho} / T}
=0 \, .
\end{eqnarray}

\begin{eqnarray}
&& \Pi_{1/2,-,\Xi_c,0,1,\rho}= f_{\Xi_c,0,1,\rho}^2 e^{-2 \overline{\Lambda}_{\Xi_c,0,1,\rho} / T}
\\ \quad && =
\int_{2m_s}^{\omega_c} [ \frac{1}{17920\pi^4}\omega^7]e^{-\omega/T}d\omega -\frac{3m_s\langle \bar q q \rangle}{8 \pi^2} T^4-\frac{3m_s\langle \bar s s \rangle}{16 \pi^2} T^4
+\frac{\langle g_s^2 GG \rangle}{512 \pi^4} T^4+\frac{\langle g_s \bar q \sigma Gq \rangle \langle \bar s s \rangle}{8}
\nonumber\\ \quad &&\quad
+\frac{\langle g_s \bar s \sigma Gs \rangle \langle \bar q q \rangle}{8}-\frac{\langle g_s \bar q \sigma Gq \rangle\langle g_s \bar s \sigma Gs \rangle}{16} {1 \over T^2} \, ,
\nonumber \\
&& f_{\Xi_c,0,1,\rho}^2 K_{\Xi_c,0,1,\rho} e^{-2 \overline{\Lambda}_{\Xi_c,0,1,\rho} / T}
\nonumber\\ \quad && =\int_{2m_s}^{\omega_c} [ -\frac{1}{80640\pi^4}\omega^9+  \frac{m_s^2}{3584\pi^4}\omega^7]e^{-\omega/T}d\omega +\frac{15m_s\langle \bar q q \rangle}{4 \pi^2} T^6-\frac{3m_s\langle \bar s s \rangle}{4\pi^2} T^6-\frac{13\langle g_s^2 GG \rangle}{ 256\pi^4}T^6
\nonumber\\ \quad &&\quad
-\frac{3m_s\langle g_s \bar q \sigma Gq \rangle}{ 4\pi^2}T^4-\frac{3m_s^2\langle g_s^2 GG \rangle}{ 256\pi^4}T^4-\frac{\langle g_s \bar q \sigma Gq \rangle\langle g_s \bar s \sigma Gs \rangle}{16}-\frac{3m_s\langle \bar q q \rangle\langle g_s^2 GG \rangle}{ 128\pi^2}T^2+\frac{m_s\langle \bar s s \rangle\langle g_s^2 GG \rangle}{ 128\pi^2}T^2
\nonumber\\ \quad &&\quad
+\frac{\langle \bar q q \rangle \langle g_s \bar s \sigma Gs \rangle\langle g_s^2 GG \rangle}{128}{1 \over T^2}+ \frac{\langle \bar s s \rangle\langle g_s \bar q \sigma Gq \rangle\langle g_s^2 GG \rangle}{128}{1 \over T^2} -\frac{\langle g_s \bar q \sigma Gq \rangle\langle g_s \bar s \sigma Gs \rangle\langle g_s^2 GG \rangle}{256}{1 \over T^4}\, ,
\nonumber \\
\nonumber && f_{\Xi_c,0,1,\rho}^2 \Sigma_{\Xi_c,0,1,\rho} e^{-2 \overline{\Lambda}_{\Xi_c,0,1,\rho} / T}
=0 \, .
\end{eqnarray}

\begin{eqnarray}
&& \Pi_{1/2,-,\Lambda_c,1,1,\rho}= f_{\Lambda_c,1,1,\rho}^2 e^{-2 \overline{\Lambda}_{\Lambda_c,1,1,\rho} / T}
\\ \quad && =\int_{0}^{\omega_c} [ \frac{3}{4480\pi^4}\omega^7]e^{-\omega/T}d\omega- \frac{\langle g_s^2 GG \rangle}{64 \pi^4} T^4 + \frac{\langle g_s \bar q \sigma Gq \rangle \langle \bar q q \rangle}{2} - \frac{\langle g_s \bar q \sigma G q\rangle^2}{8} {1 \over T^2} \, ,
\nonumber \\
&& f_{\Lambda_c,1,1,\rho}^2 K_{\Lambda_c,1,1,\rho} e^{-2 \overline{\Lambda}_{\Lambda_c,1,1,\rho} / T}
\nonumber\\ \quad && =\int_{0}^{\omega_c} [ -\frac{1}{6720\pi^4}\omega^9]e^{-\omega/T}d\omega+ \frac{9\langle g_s^2 GG \rangle}{64 \pi^4} T^6- \frac{\langle g_s \bar q \sigma G q\rangle^2}{8}+\frac{\langle \bar q q \rangle\langle g_s \bar q \sigma Gq \rangle \langle g_s^2 GG \rangle}{32}{1 \over T^2}
\nonumber\\ \quad &&\quad
- \frac{\langle g_s \bar q \sigma Gq \rangle ^2\langle g_s^2 GG \rangle}{128}{1 \over T^4} \, ,
\nonumber \\
\nonumber && f_{\Lambda_c,1,1,\rho}^2 \Sigma_{\Lambda_c,1,1,\rho} e^{-2 \overline{\Lambda}_{\Lambda_c,1,1,\rho} / T}
={3{\langle g_s^2 GG \rangle} \over 32 \pi^4} T^6 \, .
\end{eqnarray}

\begin{eqnarray}
&& \Pi_{3/2,-,\Lambda_c,2,1,\rho}= f_{\Lambda_c,2,1,\rho}^{2} e^{-2 \overline{\Lambda}_{\Lambda_c,2,1,\rho} / T}
\\ \quad && =\int_{0}^{\omega_c} [ \frac{1}{2016\pi^4}\omega^7]e^{-\omega/T}d\omega  - \frac{65\langle g_s^2 GG \rangle}{576 \pi^4} T^4 + \frac{5\langle g_s \bar q \sigma Gq \rangle\langle \bar q q \rangle}{9}-\frac{5\langle g_s \bar q \sigma Gq \rangle^2}{36} {1 \over T^2} \, ,
\nonumber \\
&& f_{\Lambda_c,2,1,\rho}^{2} K_{\Lambda_c,2,1,\rho} e^{-2 \overline{\Lambda}_{\Lambda_c,2,1,\rho} / T}
\nonumber\\ \quad && =\int_{0}^{\omega_c} [ -\frac{41}{362880\pi^4}\omega^9]e^{-\omega/T}d\omega + \frac{1069\langle g_s^2 GG \rangle}{1152\pi^4} T^6-\frac{\langle g_s \bar q \sigma Gq \rangle^2}{9}
+\frac{5\langle \bar q q \rangle\langle g_s \bar q \sigma Gq \rangle\langle g_s^2 GG \rangle}{144}{1 \over T^2}
\nonumber\\ \quad &&\quad
-\frac{5\langle g_s \bar q \sigma Gq \rangle^2\langle g_s^2 GG \rangle}{576}{1 \over T^4}\, ,
\nonumber \\
\nonumber && f_{\Lambda_c,2,1,\rho}^{2} \Sigma_{\Lambda_c,2,1,\rho} e^{-2 \overline{\Lambda}_{\Lambda_c,2,1,\rho} / T}
=\frac{5\langle g_s^2 GG \rangle}{48 \pi^4} T^6\, .
\end{eqnarray}

\begin{eqnarray}
&& \Pi_{3/2,-,\Xi_c,2,1,\rho}= f_{\Xi_c,2,1,\rho}^{2} e^{-2 \overline{\Lambda}_{\Xi_c,2,1,\rho} / T}
\\ \quad && =\int_{2m_s}^{\omega_c} [ \frac{1}{2016\pi^4}\omega^7-\frac{m_s^2}{64\pi^4}\omega^5]e^{-\omega/T}d\omega  - \frac{5m_s\langle \bar qq \rangle}{6 \pi^2} T^4 +\frac{5m_s\langle \bar s s \rangle}{6 \pi^2} T^4- \frac{65\langle g_s^2 GG \rangle}{576\pi^4} T^4+\frac{5m_s^2\langle g_s^2 GG \rangle}{128 \pi^4} T^2
\nonumber\\ \quad &&\quad
+\frac{5\langle g_s \bar q \sigma Gq \rangle\langle \bar s s \rangle}{18}+\frac{5\langle g_s \bar s \sigma Gs \rangle\langle \bar q q \rangle}{18}-\frac{5m_s\langle \bar s s \rangle\langle g_s^2 GG \rangle}{192 \pi^2}-\frac{5\langle g_s \bar q \sigma Gq \rangle\langle g_s \bar s \sigma Gs \rangle}{36} {1 \over T^2} \, ,
\nonumber \\
&& f_{\Xi_c,2,1,\rho}^{2} K_{\Xi_c,2,1,\rho} e^{-2 \overline{\Lambda}_{\Xi_c,2,1,\rho} / T}
\nonumber\\ \quad && =\int_{2m_s}^{\omega_c} [ -\frac{41}{362880\pi^4}\omega^9+\frac{53m_s^2}{10080\pi^4}\omega^7]e^{-\omega/T}d\omega + \frac{26m_s\langle \bar q q \rangle}{3\pi^2} T^6-\frac{37m_s\langle \bar s s \rangle}{3\pi^2} T^6+\frac{1069\langle g_s^2 GG \rangle}{1152\pi^4} T^6
\nonumber\\ \quad &&\quad
-\frac{11m_s\langle g_s \bar q \sigma Gq \rangle}{6\pi^2} T^4
-\frac{89m_s^2\langle g_s^2 GG \rangle}{384\pi^4} T^4
-\frac{\langle g_s \bar q \sigma Gq \rangle\langle g_s \bar s \sigma Gs \rangle}{9}-\frac{5m_s\langle \bar q q \rangle\langle g_s^2 GG \rangle}{96\pi^2} T^2+\frac{11m_s\langle \bar s s \rangle\langle g_s^2 GG \rangle}{144\pi^2} T^2
\nonumber\\ \quad &&\quad
+\frac{5\langle \bar q q \rangle\langle g_s \bar s \sigma Gs \rangle\langle g_s^2 GG \rangle}{288}{1 \over T^2}
+\frac{5\langle \bar s s \rangle\langle g_s \bar q \sigma Gq \rangle\langle g_s^2 GG \rangle}{288}{1 \over T^2}-\frac{5\langle g_s \bar q \sigma Gq \rangle\langle g_s \bar s \sigma Gs \rangle\langle g_s^2 GG \rangle}{576}{1 \over T^4}\, ,
\nonumber \\
\nonumber && f_{\Xi_c,2,1,\rho}^{2} \Sigma_{\Xi_c,2,1,\rho} e^{-2 \overline{\Lambda}_{\Xi_c,2,1,\rho} / T}
=\frac{5\langle g_s^2 GG \rangle}{48 \pi^4} T^6-\frac{5m_s^2\langle g_s^2 GG \rangle}{128 \pi^4} T^4+\frac{5m_s\langle \bar s s \rangle\langle g_s^2 GG \rangle}{288 \pi^2} T^2\, .
\end{eqnarray}

\begin{eqnarray}
&& \Pi_{1/2,-,\Lambda_c,1,0,\lambda}= f_{\Lambda_c,1,0,\lambda}^2 e^{-2 \overline{\Lambda}_{\Lambda_c,1,0,\lambda} / T}
\\ \quad && =
\int_{0}^{\omega_c} [ \frac{3}{17920\pi^4}\omega^7 ]e^{-\omega/T}d\omega - \frac{\langle g_s^2 GG \rangle}{512 \pi^4} T^4 + \frac{\langle g_s \bar q \sigma Gq \rangle \langle \bar q q \rangle}{4} - \frac{\langle g_s \bar q \sigma G q\rangle^2}{16} {1 \over T^2} \, ,
\nonumber \\
&& f_{\Lambda_c,1,0,\lambda}^2 K_{\Lambda_c,1,0,\lambda} e^{-2 \overline{\Lambda}_{\Lambda_c,1,0,\lambda} / T}
\nonumber\\ \quad && = \int_{0}^{\omega_c} [-{1 \over 20160 \pi^4} \omega^9] e^{-\omega/T} d\omega - \frac{\langle g_s^2 GG \rangle}{32 \pi^4} T^6- \frac{5\langle g_s \bar q \sigma G q\rangle^2}{16}
+\frac{\langle g_s \bar q \sigma Gq \rangle \langle \bar q q \rangle\langle g_s^2 GG \rangle}{192}{1 \over T^2}
\nonumber\\ \quad &&\quad
- \frac{\langle g_s \bar q \sigma Gq \rangle ^2\langle g_s^2 GG \rangle}{768}{1 \over T^4} \, ,
\nonumber \\
\nonumber && f_{\Lambda_c,1,0,\lambda}^2 \Sigma_{\Lambda_c,1,0,\lambda} e^{-2 \overline{\Lambda}_{\Lambda_c,1,0,\lambda} / T}
={{\langle g_s^2 GG \rangle}\over 64 \pi^4} T^6 \, .
\end{eqnarray}

\begin{eqnarray}
&& \Pi_{1/2,-,\Xi_c,1,0,\lambda}(\omega_c, T) = f_{\Xi_c,1,0,\lambda}^2 e^{-2 \overline{\Lambda}_{\Xi_c,1,0,\lambda} / T}
\\ \quad && =
\int_{2m_s}^{\omega_c} [ \frac{3}{17920\pi^4}\omega^7-\frac{3m_s^2}{640\pi^4}\omega^5 ]e^{-\omega/T}d\omega - \frac{3m_s\langle \bar q q \rangle}{8 \pi^2} T^4
+ \frac{3m_s\langle \bar s s \rangle}{16 \pi^2} T^4 -\frac{\langle g_s^2 GG \rangle}{512 \pi^4} T^4
\nonumber\\ \quad &&\quad
+\frac{m_s^2\langle g_s^2 GG \rangle}{256 \pi^4} T^2+\frac{\langle g_s\bar q \sigma Gq \rangle\langle \bar s s \rangle}{8}
+\frac{\langle g_s\bar s \sigma Gs \rangle\langle \bar q q \rangle}{8}-\frac{m_s\langle \bar s s \rangle\langle g_s^2 GG \rangle}{384 \pi^2}- \frac{\langle g_s \bar q \sigma G q\rangle\langle g_s\bar s \sigma Gs \rangle}{16} {1 \over T^2}
\nonumber \, ,
\nonumber \\
\nonumber && f_{\Xi_c,1,0,\lambda}^2 K_{\Xi_c,1,0,\lambda} e^{-2 \overline{\Lambda}_{\Xi_c,1,0,\lambda} / T}
\\ \quad && =\int_{2m_s}^{\omega_c} [ -\frac{1}{20160\pi^4}\omega^9+\frac{33m_s^2}{17920\pi^4}\omega^7 ]e^{-\omega/T}d\omega +\frac{15m_s\langle \bar q q \rangle}{4 \pi^2} T^6-\frac{9m_s\langle \bar s s \rangle}{4 \pi^2} T^6-\frac{\langle g_s^2 GG \rangle}{32 \pi^4} T^6
\nonumber\\ \quad &&\quad
-\frac{m_s^2\langle g_s^2 GG \rangle}{64 \pi^4} T^4-\frac{5\langle g_s \bar q \sigma Gq \rangle\langle g_s \bar s \sigma Gs \rangle}{16}-\frac{m_s\langle \bar q q \rangle\langle g_s^2 GG \rangle}{128 \pi^2} T^2+\frac{m_s\langle \bar s s \rangle\langle g_s^2 GG \rangle}{384 \pi^2} T^2
\nonumber\\ \quad &&\quad
+\frac{\langle \bar q q \rangle \langle g_s \bar s \sigma Gs \rangle\langle g_s^2 GG \rangle}{384}{1 \over T^2}+\frac{\langle g_s \bar q \sigma Gq \rangle\langle \bar s s \rangle\langle g_s^2 GG \rangle}{384}{1 \over T^2}-\frac{\langle g_s \bar q \sigma Gq \rangle\langle g_s \bar s \sigma Gs \rangle\langle g_s^2 GG \rangle}{768}{1 \over T^4} \, ,
\nonumber \\
\nonumber && f_{\Xi_c,1,0,\lambda}^2 \Sigma_{\Xi_c,1,0,\lambda} e^{-2 \overline{\Lambda}_{\Xi_c,1,0,\lambda} / T}
=\frac{\langle g_s^2 GG \rangle}{64 \pi^4} T^6-\frac{m_s^2\langle g_s^2 GG \rangle}{128 \pi^4} T^4+\frac{m_s\langle \bar s s \rangle\langle g_s^2 GG \rangle}{192 \pi^2} T^2 \, .
\end{eqnarray}

\begin{eqnarray}
&& \Pi_{1/2,-,\Sigma_c,0,1,\lambda}= f_{\Sigma_c,0,1,\lambda}^2 e^{-2 \overline{\Lambda}_{\Sigma_c,0,1,\lambda} / T}
\\ \quad && =
\int_{0}^{\omega_c} [ \frac{1}{2560\pi^4}\omega^7 ]e^{-\omega/T}d\omega +\frac{3\langle g_s^2 GG \rangle}{512 \pi^4} T^4+\frac{\langle g_s \bar q \sigma Gq \rangle \langle \bar q q \rangle}{4}-\frac{\langle g_s \bar q \sigma Gq \rangle^2}{16} {1 \over T^2} \, ,
\nonumber \\
&& f_{\Sigma_c,0,1,\lambda}^2 K_{\Sigma_c,0,1,\lambda} e^{-2 \overline{\Lambda}_{\Sigma_c,0,1,\lambda} / T}
\nonumber\\ \quad && = \int_{0}^{\omega_c} [-{1 \over 7680 \pi^4} \omega^9] e^{-\omega/T} d\omega - \frac{11\langle g_s^2 GG \rangle}{128 \pi^4} T^6-\frac{5\langle g_s \bar q \sigma Gq \rangle^2}{16}
+\frac{\langle \bar q q \rangle \langle g_s \bar q \sigma Gq \rangle\langle g_s^2 GG \rangle}{192}{1 \over T^2}
\nonumber\\ \quad &&\quad
- \frac{\langle g_s \bar q \sigma Gq \rangle^2\langle g_s^2 GG \rangle}{768}{1 \over T^4} \, ,
\nonumber \\
\nonumber && f_{\Sigma_c,0,1,\lambda}^2 \Sigma_{\Sigma_c,0,1,\lambda} e^{-2 \overline{\Lambda}_{\Sigma_c,0,1,\lambda} / T}
=0 \, .
\end{eqnarray}

\begin{eqnarray}
&& \Pi_{1/2,-,\Xi^\prime_c,0,1,\lambda}= f_{\Xi^\prime_c,0,1,\lambda}^2 e^{-2 \overline{\Lambda}_{\Xi^\prime_c,0,1,\lambda} / T}
\\ \quad && =
\int_{2m_s}^{\omega_c} [ \frac{1}{2560\pi^4}\omega^7-  \frac{9m_s^2}{640\pi^4}\omega^5]e^{-\omega/T}d\omega -\frac{3m_s\langle \bar q q \rangle}{8 \pi^2} T^4+\frac{15m_s\langle \bar s s \rangle}{16 \pi^2} T^4
\nonumber\\ \quad &&\quad
+\frac{3\langle g_s^2 GG \rangle}{512 \pi^4} T^4+\frac{\langle g_s \bar q \sigma Gq \rangle \langle \bar s s \rangle}{8}+\frac{\langle g_s \bar s \sigma Gs \rangle \langle \bar q q \rangle}{8}-\frac{\langle g_s \bar q \sigma Gq \rangle\langle g_s \bar s \sigma Gs \rangle}{16} {1 \over T^2} \, ,
\nonumber \\
&& f_{\Xi^\prime_c,0,1,\lambda}^2 K_{\Xi^\prime_c,0,1,\lambda} e^{-2 \overline{\Lambda}_{\Xi^\prime_c,0,1,\lambda} / T}
\nonumber\\ \quad && =\int_{2m_s}^{\omega_c} [ -\frac{1}{7680\pi^4}\omega^9+  \frac{111m_s^2}{17920\pi^4}\omega^7]e^{-\omega/T}d\omega +\frac{15m_s\langle \bar q q \rangle}{4 \pi^2} T^6-\frac{12m_s\langle \bar s s \rangle}{ \pi^2} T^6-\frac{11\langle g_s^2 GG \rangle}{ 128\pi^4}T^6
\nonumber\\ \quad &&\quad
-\frac{m_s^2\langle g_s^2 GG \rangle}{ 256\pi^4}T^4-\frac{5\langle g_s \bar q \sigma Gq \rangle\langle g_s \bar s \sigma Gs \rangle}{16}-\frac{m_s\langle \bar q q \rangle\langle g_s^2 GG \rangle}{ 128\pi^2}T^2+\frac{m_s\langle \bar s s \rangle\langle g_s^2 GG \rangle}{ 384\pi^2}T^2
\nonumber\\ \quad &&\quad
+\frac{\langle \bar q q \rangle \langle g_s \bar s \sigma Gs \rangle\langle g_s^2 GG \rangle}{384}{1 \over T^2}+ \frac{\langle \bar s s \rangle\langle g_s \bar q \sigma Gq \rangle\langle g_s^2 GG \rangle}{384}{1 \over T^2} -\frac{\langle g_s \bar q \sigma Gq \rangle\langle g_s \bar s \sigma Gs \rangle\langle g_s^2 GG \rangle}{768}{1 \over T^4}\, ,
\nonumber \\
\nonumber && f_{\Xi^\prime_c,0,1,\lambda}^2 \Sigma_{\Xi^\prime_c,0,1,\lambda} e^{-2 \overline{\Lambda}_{\Xi^\prime_c,0,1,\lambda} / T}
=0 \, .
\end{eqnarray}

\begin{eqnarray}
&& \Pi_{1/2,-,\Omega_c,0,1,\lambda}= f_{\Omega_c,0,1,\lambda}^2 e^{-2 \overline{\Lambda}_{\Omega_c,0,1,\lambda} / T}
\\ \quad && =
\int_{4m_s}^{\omega_c} [ \frac{1}{2560\pi^4}\omega^7-  \frac{3m_s^2}{128\pi^4}\omega^5+\frac{15m_s^4}{64\pi^4}\omega^3]e^{-\omega/T}d\omega +\frac{9m_s\langle \bar s s \rangle}{8 \pi^2} T^4+\frac{3\langle g_s^2 GG \rangle}{512 \pi^4} T^4
\nonumber\\ \quad &&\quad
-\frac{3m_s^3\langle \bar s s \rangle}{2 \pi^2} T^2+\frac{\langle g_s \bar s \sigma Gs \rangle \langle \bar s s \rangle}{4}+\frac{3m_s^2\langle \bar s s \rangle^2}{8}-\frac{\langle g_s \bar s \sigma Gs \rangle^2}{16} {1 \over T^2} \, ,
\nonumber \\
&& f_{\Omega_c,0,1,\lambda}^2 K_{\Omega_c,0,1,\lambda} e^{-2 \overline{\Lambda}_{\Omega_c,0,1,\lambda} / T}
\nonumber\\ \quad && =\int_{4m_s}^{\omega_c} [ -\frac{1}{7680\pi^4}\omega^9+  \frac{3m_s^2}{280\pi^4}\omega^7]e^{-\omega/T}d\omega -\frac{33m_s\langle \bar s s \rangle}{2 \pi^2} T^6-\frac{11\langle g_s^2 GG \rangle}{ 128\pi^4} T^6+\frac{m_s^2\langle g_s^2 GG \rangle}{ 256\pi^4}T^4-\frac{5\langle g_s \bar s \sigma Gs \rangle^2}{16}
\nonumber\\ \quad &&\quad
-\frac{m_s\langle \bar s s \rangle\langle g_s^2 GG \rangle}{ 96\pi^2}T^2+\frac{\langle \bar s s \rangle\langle g_s \bar s \sigma Gs \rangle\langle g_s^2 GG \rangle}{192}{1 \over T^2}-\frac{m_s^2\langle \bar s s \rangle^2\langle g_s^2 GG \rangle}{1152}{1 \over T^2}-\frac{\langle g_s \bar s \sigma Gs \rangle^2\langle g_s^2 GG \rangle}{768}{1 \over T^4}\, ,
\nonumber \\
\nonumber && f_{\Omega_c,0,1,\lambda}^2 \Sigma_{\Omega_c,0,1,\lambda} e^{-2 \overline{\Lambda}_{\Omega_c,0,1,\lambda} / T}
=0 \, .
\end{eqnarray}

\begin{eqnarray}
&& \Pi_{1/2,-,\Sigma_c,1,1,\lambda}= f_{\Sigma_c,1,1,\lambda}^2 e^{-2 \overline{\Lambda}_{\Sigma_c,1,1,\lambda} / T}
\\ \quad && =
\int_{0}^{\omega_c} [ {1 \over 4480 \pi^4} \omega^7 ] e^{-\omega/T}d\omega - \frac{\langle g_s^2 GG \rangle}{64 \pi^4} T^4 + \frac{\langle g_s \bar q \sigma Gq \rangle \langle \bar q q \rangle}{2} - \frac{\langle g_s \bar q \sigma G q\rangle^2}{8} {1 \over T^2} \, ,
\nonumber \\
&& f_{\Sigma_c,1,1,\lambda}^2 K_{\Sigma_c,1,1,\lambda} e^{-2 \overline{\Lambda}_{\Sigma_c,1,1,\lambda} / T}
\nonumber\\ \quad && = \int_{0}^{\omega_c} [-{1 \over 13440 \pi^4} \omega^9] e^{-\omega/T} d\omega + \frac{3\langle g_s^2 GG \rangle}{32 \pi^4} T^6- \frac{5\langle g_s \bar q \sigma G q\rangle^2}{8}
+\frac{\langle \bar q q \rangle\langle g_s \bar q \sigma Gq \rangle \langle g_s^2 GG \rangle}{96}{1 \over T^2}
\nonumber\\ \quad &&\quad
- \frac{\langle g_s \bar q \sigma Gq \rangle ^2\langle g_s^2 GG \rangle}{384}{1 \over T^4} \, ,
\nonumber \\
\nonumber && f_{\Sigma_c,1,1,\lambda}^2 \Sigma_{\Sigma_c,1,1,\lambda} e^{-2 \overline{\Lambda}_{\Sigma_c,1,1,\lambda} / T}
={{\langle g_s^2 GG \rangle}\over 32 \pi^4} T^6 \, .
\end{eqnarray}

\begin{eqnarray}
&& \Pi_{1/2,-,\Xi^\prime_c,1,1,\lambda}= f_{\Xi^\prime_c,1,1,\lambda}^2 e^{-2 \overline{\Lambda}_{\Xi^\prime_c,1,1,\lambda} / T}
\\ \quad && =\int_{2m_s}^{\omega_c} [ \frac{1}{4480\pi^4}\omega^7-  \frac{3m_s^2}{640\pi^4}\omega^5]e^{-\omega/T}d\omega  - \frac{3m_s\langle \bar q q \rangle}{4 \pi^2} T^4 - \frac{\langle g_s^2 GG \rangle}{64 \pi^4} T^4 +\frac{m_s^2\langle g_s^2 GG \rangle}{256 \pi^4} T^2
\nonumber\\ \quad &&\quad
+ \frac{\langle g_s \bar q \sigma Gq \rangle\langle \bar s s \rangle}{4}+ \frac{\langle g_s \bar s \sigma Gs \rangle\langle \bar q q \rangle}{4} -\frac{m_s\langle \bar s s \rangle\langle g_s^2 GG \rangle}{384 \pi^2}-\frac{\langle g_s \bar q \sigma Gq \rangle\langle g_s \bar s \sigma Gs \rangle}{8}{1 \over T^2} \, ,
\nonumber \\
&& f_{\Xi^\prime_c,1,1,\lambda}^2 K_{\Xi^\prime_c,1,1,\lambda} e^{-2 \overline{\Lambda}_{\Xi^\prime_c,1,1,\lambda} / T}
\nonumber\\ \quad && =\int_{2m_s}^{\omega_c} [ -\frac{1}{13440\pi^4}\omega^9+  \frac{3m_s^2}{1280\pi^4}\omega^7]e^{-\omega/T}d\omega + \frac{15m_s\langle \bar q q \rangle}{2 \pi^2} T^6- \frac{3m_s\langle \bar s s \rangle}{2 \pi^2} T^6+\frac{3\langle g_s^2 GG \rangle}{32 \pi^4} T^6-\frac{7m_s^2\langle g_s^2 GG \rangle}{256 \pi^4} T^4
\nonumber\\ \quad &&\quad
-\frac{5\langle g_s \bar q \sigma G q\rangle\langle g_s \bar s \sigma Gs \rangle}{8}-\frac{m_s\langle \bar q q \rangle\langle g_s^2 GG \rangle}{64 \pi^2} T^2+\frac{m_s\langle \bar s s \rangle\langle g_s^2 GG \rangle}{96\pi^2} T^2+\frac{\langle \bar q q \rangle\langle g_s \bar s \sigma Gs \rangle\langle g_s^2 GG \rangle}{192}{1 \over T^2}
\nonumber\\ \quad &&\quad
+\frac{\langle \bar s s \rangle\langle g_s \bar q \sigma Gq \rangle\langle g_s^2 GG \rangle}{192}{1 \over T^2}-\frac{\langle g_s \bar q \sigma Gq \rangle\langle g_s \bar s \sigma Gs \rangle\langle g_s^2 GG \rangle}{384}{1 \over T^4}\, ,
\nonumber \\
\nonumber && f_{\Xi^\prime_c,1,1,\lambda}^2 \Sigma_{\Xi^\prime_c,1,1,\lambda} e^{-2 \overline{\Lambda}_{\Xi^\prime_c,1,1,\lambda} / T}
=\frac{\langle g_s^2 GG \rangle}{32 \pi^4} T^6- \frac{m_s^2\langle g_s^2 GG \rangle}{128 \pi^4} T^4\, .
\end{eqnarray}

\begin{eqnarray}
&& \Pi_{1/2,-,\Omega_c,1,1,\lambda}= f_{\Omega_c,1,1,\lambda}^2 e^{-2 \overline{\Lambda}_{\Omega_c,1,1,\lambda} / T}
\\ \quad && =\int_{4m_s}^{\omega_c} [ \frac{1}{4480\pi^4}\omega^7]e^{-\omega/T}d\omega  - \frac{3m_s\langle \bar s s \rangle}{2 \pi^2} T^4 - \frac{\langle g_s^2 GG \rangle}{64 \pi^4} T^4 +\frac{3m_s^3\langle \bar s s \rangle}{4 \pi^2} T^2+\frac{m_s^2\langle g_s^2 GG \rangle}{128\pi^4} T^2
\nonumber\\ \quad &&\quad
+ \frac{\langle g_s \bar s \sigma Gs \rangle\langle \bar s s \rangle}{2}- \frac{m_s^2\langle \bar s s \rangle^2}{2} -\frac{m_s\langle \bar s s \rangle\langle g_s^2 GG \rangle}{192 \pi^2}-\frac{\langle g_s \bar s \sigma Gs \rangle^2}{8}{1 \over T^2} \, ,
\nonumber \\
&& f_{\Omega_c,1,1,\lambda}^2 K_{\Omega_c,1,1,\lambda} e^{-2 \overline{\Lambda}_{\Omega_c,1,1,\lambda} / T}
\nonumber\\ \quad && =\int_{4m_s}^{\omega_c} [ -\frac{1}{13440\pi^4}\omega^9+  \frac{3m_s^2}{2240\pi^4}\omega^7]e^{-\omega/T}d\omega + \frac{12m_s\langle \bar s s \rangle}{\pi^2} T^6+\frac{3\langle g_s^2 GG \rangle}{32 \pi^4}T^6-\frac{m_s^2\langle g_s^2 GG \rangle}{32 \pi^4}T^4-\frac{5\langle g_s \bar s \sigma Gs \rangle^2}{8}
\nonumber\\ \quad &&\quad
-\frac{m_s\langle \bar s s \rangle\langle g_s^2 GG \rangle}{96\pi^2}T^2+\frac{\langle \bar s s \rangle\langle g_s \bar s \sigma Gs \rangle\langle g_s^2 GG \rangle}{96}{1 \over T^2}-\frac{m_s^2\langle \bar s s \rangle^2\langle g_s^2 GG \rangle}{576}{1 \over T^2}-\frac{\langle g_s \bar s \sigma Gs \rangle^2\langle g_s^2 GG \rangle}{384}{1 \over T^4}\, ,
\nonumber \\
\nonumber && f_{\Omega_c,1,1,\lambda}^2 \Sigma_{\Omega_c,1,1,\lambda} e^{-2 \overline{\Lambda}_{\Omega_c,1,1,\lambda} / T}
=\frac{\langle g_s^2 GG \rangle}{32 \pi^4} T^6- \frac{m_s^2\langle g_s^2 GG \rangle}{64\pi^4} T^4\, .
\end{eqnarray}

\begin{eqnarray}
&& \Pi_{3/2,-,\Sigma_c,2,1,\lambda}= f_{\Sigma_c,2,1,\lambda}^{2} e^{-2 \overline{\Lambda}_{\Sigma_c,2,1,\lambda} / T}
\\ \quad && =\int_{0}^{\omega_c} [ \frac{1}{2016\pi^4}\omega^7]e^{-\omega/T}d\omega  - \frac{5\langle g_s^2 GG \rangle}{192 \pi^4} T^4 + \frac{5\langle g_s \bar q \sigma Gq \rangle\langle \bar q q \rangle}{9}-\frac{5\langle g_s \bar q \sigma Gq \rangle^2}{36} {1 \over T^2} \, ,
\nonumber \\
&& f_{\Sigma_c,2,1,\lambda}^{2} K_{\Sigma_c,2,1,\lambda} e^{-2 \overline{\Lambda}_{\Sigma_c,2,1,\lambda} / T}
\nonumber\\ \quad && =\int_{0}^{\omega_c} [ -\frac{1}{5760\pi^4}\omega^9]e^{-\omega/T}d\omega + \frac{205\langle g_s^2 GG \rangle}{1152\pi^4} T^6-\frac{13\langle g_s \bar q \sigma Gq \rangle^2}{18}
+\frac{5\langle \bar q q \rangle\langle g_s \bar q \sigma Gq \rangle\langle g_s^2 GG \rangle}{432}{1 \over T^2}
\nonumber\\ \quad &&\quad
-\frac{5\langle g_s \bar q \sigma Gq \rangle^2\langle g_s^2 GG \rangle}{1728}{1 \over T^4}\, ,
\nonumber \\
\nonumber && f_{\Sigma_c,2,1,\lambda}^{2} \Sigma_{\Sigma_c,2,1,\lambda} e^{-2 \overline{\Lambda}_{\Sigma_c,2,1,\lambda} / T}
=\frac{5\langle g_s^2 GG \rangle}{48 \pi^4} T^6\, .
\end{eqnarray}

\begin{eqnarray}
&& \Pi_{3/2,-,\Xi^\prime_c,2,1,\lambda}= f_{\Xi^\prime_c,2,1,\lambda}^{2} e^{-2 \overline{\Lambda}_{\Xi^\prime_c,2,1,\lambda} / T}
\\ \quad && =\int_{2m_s}^{\omega_c} [ \frac{1}{2016\pi^4}\omega^7-\frac{m_s^2}{64\pi^4}\omega^5]e^{-\omega/T}d\omega  - \frac{5m_s\langle \bar qq \rangle}{6 \pi^2} T^4 +\frac{5m_s\langle \bar s s \rangle}{6 \pi^2} T^4- \frac{5\langle g_s^2 GG \rangle}{192\pi^4} T^4+\frac{5m_s^2\langle g_s^2 GG \rangle}{384 \pi^4} T^2
\nonumber\\ \quad &&\quad
+\frac{5\langle g_s \bar q \sigma Gq \rangle\langle \bar s s \rangle}{18}+\frac{5\langle g_s \bar s \sigma Gs \rangle\langle \bar q q \rangle}{18}-\frac{5m_s\langle \bar s s \rangle\langle g_s^2 GG \rangle}{576 \pi^2}-\frac{5\langle g_s \bar q \sigma Gq \rangle\langle g_s \bar s \sigma Gs \rangle}{36} {1 \over T^2} \, ,
\nonumber \\
&& f_{\Xi^\prime_c,2,1,\lambda}^{2} K_{\Xi^\prime_c,2,1,\lambda} e^{-2 \overline{\Lambda}_{\Xi^\prime_c,2,1,\lambda} / T}
\nonumber\\ \quad && =\int_{2m_s}^{\omega_c} [ -\frac{1}{5760\pi^4}\omega^9+\frac{5m_s^2}{672\pi^4}\omega^7]e^{-\omega/T}d\omega + \frac{26m_s\langle \bar q q \rangle}{3\pi^2} T^6-\frac{37m_s\langle \bar s s \rangle}{3\pi^2} T^6+\frac{205\langle g_s^2 GG \rangle}{1152\pi^4} T^6-\frac{89m_s^2\langle g_s^2 GG \rangle}{1152\pi^4} T^4
\nonumber\\ \quad &&\quad
-\frac{13\langle g_s \bar q \sigma Gq \rangle\langle g_s \bar s \sigma Gs \rangle}{18}-\frac{5m_s\langle \bar q q \rangle\langle g_s^2 GG \rangle}{288\pi^2} T^2+\frac{11m_s\langle \bar s s \rangle\langle g_s^2 GG \rangle}{432\pi^2} T^2+\frac{5\langle \bar q q \rangle\langle g_s \bar s \sigma Gs \rangle\langle g_s^2 GG \rangle}{864}{1 \over T^2}
\nonumber\\ \quad &&\quad
+\frac{5\langle \bar s s \rangle\langle g_s \bar q \sigma Gq \rangle\langle g_s^2 GG \rangle}{864}{1 \over T^2}-\frac{5\langle g_s \bar q \sigma Gq \rangle\langle g_s \bar s \sigma Gs \rangle\langle g_s^2 GG \rangle}{1728}{1 \over T^4}\, ,
\nonumber \\
\nonumber && f_{\Xi^\prime_c,2,1,\lambda}^{2} \Sigma_{\Xi^\prime_c,2,1,\lambda} e^{-2 \overline{\Lambda}_{\Xi^\prime_c,2,1,\lambda} / T}
=\frac{5\langle g_s^2 GG \rangle}{48 \pi^4} T^6-\frac{5m_s^2\langle g_s^2 GG \rangle}{128 \pi^4} T^4+\frac{5m_s\langle \bar s s \rangle\langle g_s^2 GG \rangle}{288 \pi^2} T^2\, .
\end{eqnarray}

\begin{eqnarray}
&& \Pi_{3/2,-,\Omega_c,2,1,\lambda}= f_{\Omega_c,2,1,\lambda}^{2} e^{-2 \overline{\Lambda}_{\Omega_c,2,1,\lambda} / T}
\\ \quad && =\int_{4m_s}^{\omega_c} [ \frac{1}{2016\pi^4}\omega^7-\frac{m_s^2}{48\pi^4}\omega^5+\frac{5m_s^4}{24\pi^4}\omega^3]e^{-\omega/T}d\omega  -\frac{5\langle g_s^2 GG \rangle}{192\pi^4} T^4-\frac{5m_s^3\langle \bar s s \rangle}{6 \pi^2} T^2 +\frac{5m_s^2\langle g_s^2 GG \rangle}{192 \pi^4} T^2
\nonumber\\ \quad &&\quad
+\frac{5\langle g_s \bar s \sigma Gs \rangle\langle \bar s s \rangle}{9}-\frac{5m_s\langle \bar s s \rangle\langle g_s^2 GG \rangle}{288 \pi^2}-\frac{5\langle g_s \bar s \sigma Gs \rangle^2}{36} {1 \over T^2} \, ,
\nonumber \\
&& f_{\Omega_c,2,1,\lambda}^{2} K_{\Omega_c,2,1,\lambda} e^{-2 \overline{\Lambda}_{\Omega_c,2,1,\lambda} / T}
\nonumber\\ \quad && =\int_{4m_s}^{\omega_c} [ -\frac{1}{5760\pi^4}\omega^9+\frac{37m_s^2}{3360\pi^4}\omega^7]e^{-\omega/T}d\omega - \frac{22m_s\langle \bar s s \rangle}{3\pi^2} T^6+\frac{205\langle g_s^2 GG \rangle}{1152\pi^4} T^6-\frac{37m_s^2\langle g_s^2 GG \rangle}{288\pi^4} T^4-\frac{13\langle g_s \bar s \sigma Gs \rangle^2}{18}
\nonumber\\ \quad &&\quad
+\frac{7m_s\langle \bar s s \rangle\langle g_s^2 GG \rangle}{432\pi^2} T^2+\frac{5\langle \bar s s \rangle\langle g_s \bar s \sigma Gs \rangle\langle g_s^2 GG \rangle}{432}{1 \over T^2}-\frac{m_s^2\langle \bar s s \rangle^2\langle g_s^2 GG \rangle}{648}{1 \over T^2}
-\frac{5\langle g_s \bar s \sigma Gs \rangle^2\langle g_s^2 GG \rangle}{1728}{1 \over T^4}\, ,
\nonumber \\
\nonumber && f_{\Omega_c,2,1,\lambda}^{2} \Sigma_{\Omega_c,2,1,\lambda} e^{-2 \overline{\Lambda}_{\Omega_c,2,1,\lambda} / T}
=\frac{5\langle g_s^2 GG \rangle}{48 \pi^4} T^6-\frac{5m_s^2\langle g_s^2 GG \rangle}{64 \pi^4} T^4+\frac{5m_s\langle \bar s s \rangle\langle g_s^2 GG \rangle}{144 \pi^2} T^2\, .
\end{eqnarray}

\end{document}